\documentclass[twocolumn,superscriptaddress,showpacs,preprintnumbers,nofootinbib,amsmath,amssymb,floatfix,aps]{revtex4-1}
\usepackage{dcolumn}
\usepackage{bm}
\usepackage{color}
\usepackage{listings} 
\usepackage{url}
\usepackage{graphicx}
\usepackage{amsmath}
\usepackage{amssymb}
\usepackage{epstopdf}
\usepackage{braket}
\usepackage{slashed}
\usepackage{bbm}
\usepackage{bbold}
\usepackage[pdftex, colorlinks=false]{hyperref}
\usepackage{enumerate}
\usepackage[normalem]{ulem}

\DeclareMathAlphabet\mathbb{U}{msb}{m}{n}

\usepackage{float} 
\extrafloats{200}
\def\be{\begin{equation}}
\def\ee{\end{equation}}
\def\bea{\begin{eqnarray}}
\def\eea{\end{eqnarray}}
\def\bear{\begin{array}}
\def\ear{\end{array}}
\def\bfig{\begin{figure}}
\def\efig{\end{figure}}
\def\bcen{\begin{center}}
\def\ecen{\end{center}}
\def\bi{\begin{itemize}}
\def\ei{\end{itemize}}

\def\raw{\rightarrow}

\makeatletter
\DeclareRobustCommand{\filename}[1]{%
 \begingroup
  \def\textendash{-}%
  \filename@parse{#1}%
  \edef\filename@base{\detokenize\expandafter{\filename@base}}%
  \texttt{\filename@base.\filename@ext}%
 \endgroup
}
\makeatother

\begin{abstract}
The Bayesian approach for feed-forward neural networks has been applied to the extraction of the nucleon axial form factor from the neutrino-deuteron scattering data measured by the Argonne National Laboratory bubble chamber experiment. This framework allows to perform a model-independent determination of the axial form factor from data. When the low $0.05 < Q^2 < 0.10$~GeV$^2$ data are included in the analysis, the resulting axial radius disagrees with available determinations. Furthermore, a large sensitivity to the corrections from the deuteron structure is obtained. In turn, when the low-$Q^2$ region is not taken into account with or without deuteron corrections, no significant deviations from previous determinations have been observed. A more accurate determination of the nucleon axial form factor requires new precise measurements of neutrino-induced quasielastic scattering on hydrogen and deuterium.    
\end{abstract}

\begin{document}
\title{Nucleon axial form factor from a Bayesian neural-network analysis of neutrino-scattering data}

\author{Luis Alvarez-Ruso}
\affiliation{Departamento de F\'isica Te\'orica and Instituto de F\'\i sica Corpuscular (IFIC), Centro Mixto UVEG-CSIC, Valencia, Spain}
\author{Krzysztof M. Graczyk}
\affiliation{Institute of Theoretical Physics, University of Wroc\l aw, pl. M. Borna 9, 50-204, Wroc\l aw, Poland}
\author{Eduardo Saul-Sala}
\affiliation{Departamento de F\'isica Te\'orica and Instituto de F\'\i sica Corpuscular (IFIC), Centro Mixto UVEG-CSIC, Valencia, Spain}


\maketitle



\section{Introduction}

A good understanding of neutrino interactions with matter are crucial to achieve the precision goals of oscillation experiments that aim at a precise determination of neutrino properties~\cite{Alvarez-Ruso:2017oui}. In current (T2K, NOvA) and future (DUNE, HyperK) oscillation experiments with few-GeV neutrinos, a realistic modeling of neutrino interactions with nuclei, and their uncertainties in a broad kinematic range is required to  distinguish signal from background and minimize systematic errors. A key ingredient of such models are the amplitudes and cross sections at the nucleon level.

In particular, a source of uncertainty arises from the nucleon axial form factor $F_A$. This fundamental nucleon property is a function of $Q^2$, defined as minus the four-momentum transferred to the nucleon squared. The axial coupling $g_A = F_A(Q^2=0)$ is known rather precisely from the neutron $\beta$-decay asymmetry~\cite{Patrignani:2016xqp}: 
\begin{equation}
\label{Eq:ga_with_error}
g_A = 1.2723 \pm 0.0023 \,,
\end{equation}
although a more precise value can be obtained using recent measurements of the nucleon lifetime~\cite{Gonzalez-Alonso:2018omy}.
For the $Q^2$ dependence, the most common parametrization is the dipole ansatz
\be
\label{Eq:FA_dipole}
F_A^{\mathrm{dipole}} (Q^2) = g_A \left( 1 + \frac{Q^2}{M_A^2} \right)^{-2} \,,
\ee
in terms of a single parameter, the so-called axial mass $M_A$. The dipole parametrization has been utilized to describe also the electric and magnetic form factors of the nucleon. In the Breit frame and for small momenta, this $Q^2$ dependence implies that the charge distribution is an exponentially decreasing function of the radial coordinate. Both the electric and the magnetic form factors of the nucleon deviate from the dipole parametrization, for a review see Ref.~\cite{Arrington:2006zm}. It seems then natural to expect similar deviations for the axial one.  

Empirical information about $F_A$ can be obtained from neutrino charged-current quasielastic (CCQE) scattering $\nu_l \, n \raw l^- \, p$. Modern neutrino cross-section measurements have been performed on heavy nuclear targets (mostly $^{12}$C) where the determination of $F_A$ becomes unreliable due to the presence of not well-constrained nuclear corrections and the difficulties in isolating the CCQE channel in a model-independent way. A detailed discussion of this problem can be found, for instance, in Sec. III of Ref.~\cite{Katori:2016yel}. A more direct and, in principle, less model dependent determination of $F_A$ relies on bubble-chamber data on deuterium. Global analyses of the Argonne National Laboratory (ANL)~\cite{Mann:1973pr,Barish:1977qk,Miller:1982qi}, Brookhaven
National Laboratory~\cite{Baker:1981su,Kitagaki:1990vs}, Fermilab~\cite{Kitagaki:1983px} and CERN~\cite{Allasia:1990uy} data with updated vector from factors based on modern electron-scattering data have been performed by Bodek and collaborators. A reference value of  $M_A = 1.016 \pm 0.026$~GeV with a small (2.5\%) error has been obtained~\cite{Bodek:2007ym}. 

On the other hand, as pointed out in Refs.~\cite{Bhattacharya:2011ah,Bhattacharya:2015mpa}, anzatz~(\ref{Eq:FA_dipole}) is not theoretically well founded. A new extraction of $F_A$ has been recently undertaken using a  functional representation of the form factor based on conformal mapping ({\it z} expansion)~\cite{Meyer:2016oeg}. The function is only constrained by the analytic structure and asymptotic behavior dictated by QCD. The resulting form factor is consistent with the dipole one but with a much larger error, Fig.~7 of Ref.~\cite{Meyer:2016oeg}. In particular, the axial radius,
\be
\label{Eq:ra2}
r_A^2 \equiv - \frac{6}{g_A} \left. \frac{d F_A}{d Q^2}\right|_{Q^2 =0} 
\ee
obtained is $r_A^2 = 0.46 \pm 0.22$~fm$^2$, which agrees with the dipole one $ r_A^2 = 12/M_A^2$ but with an $\sim 20$ times larger error. This result might have implications for oscillation studies and calls for a new measurement of neutrino-nucleon cross sections, which is in any case desirable. The axial radius can also be extracted from muon capture by protons. A recent analysis~\cite{Hill:2017wgb} using the {\it z} expansion obtains $r_A^2 = 0.43 \pm 0.24$~fm$^2$, in agreement with the neutrino-scattering result.    

A promising source of information about $F_A(Q^2)$ is lattice QCD. Although the experimental value of $g_A$ has been recurrently underpredicted in lattice QCD studies, the use of improved algorithms has recently lead to consistent results~\cite{Berkowitz:2017gql,Alexandrou:2017hac,Capitani:2017qpc,Rajan:2017lxk}. A global analysis of the low-$Q^2$ and light-quark mass dependence of the results of Refs.~\cite{Alexandrou:2017hac,Capitani:2017qpc,Rajan:2017lxk} using baryon chiral perturbation theory has found $g_A = 1.237 \pm 0.074$ and $r_A^2 = 0.263 \pm 0.038$~\cite{Yao:2017fym}. The central value of $r_A^2$ is considerably lower than those from empirical determinations but within the (large) error bars of the {\it z} expansion results.      

The choice of a specific functional form of $F_A$ may bias the results of the analysis. Moreover, the choice of the number of parameters within a given parametrization is a delicate question. Too few parameters may not give enough versatility. As the number of parameters increase, the $\chi^2$ value of the fits can be reduced, but at some point the fit tends to reproduce statistical fluctuations of the experimental data~\cite{Graczyk:2014lba}. A reduction of the model-dependence of the results can be obtained within the methods of neural networks.  This approach has been used to obtain nucleon parton distribution functions from deep-inelastic scattering data by the neural network parton
distribution function (NNPDF) collaboration \cite{Ball:2013lla}. 

In this paper, we demonstrate that  model-independent information about $F_A$ can be obtained from a semi-parametric analysis of $\nu$-deuteron scattering data\footnote{See Sec.~\ref{Sec:Bayesian:sub:Feed-Forward} for more detailed descriptions of parametric and semiparametric techniques.}. 
 In contrast to the parametric approach in which a particular parametrization of $F_A$ is adopted based on physics assumptions, semi-parametric ones are not motivated by physics; they allow to construct a statistical model in terms of an ensemble of probability densities that are used to do statistical inference {\it i.e.} to determine the quantities of interest and their uncertainties (see Chap. 2 of Ref.~\cite{Bishop_book}).
The lack of physics motivation may prevent the results from being extrapolated outside the fit region (positive $Q^2$ in our case). 
On the other hand, given the generality of the approach, the results may contain new physics beyond the underlying
assumptions of a given model or be affected by theoretical mismodeling and/or deficiencies in a data set.

To perform this semiparametric analysis, we make use of feed-forward neural networks\footnote{Semiparametric analyses of experimental data can also rely on other families of functions, such as polynomials or radial-basis functions~\cite{Bishop_book}.}, a class of functions with unlimited adaptive abilities~\cite{Hornik89}.  With this choice, we can eliminate any bias in the result introduced by the functional form of the fit function. Depending on the number of adaptive parameters, one can get different variants of the statistical model (fit). In this context, Bayesian statistics has proved to be a very effective tool~\cite{DAgostini_book}. Its methods allow to make comparisons between different models and control the number of parameters in the fit. Bayesian methods are successfully used in different branches of physics. For instance, in hadron and nuclear physics, they have been applied to the study of the resonance content of the $p (\gamma, K^+) \Lambda$ reaction~\cite{DeCruz:2011xi} and to constrain the nuclear energy-density functional from nuclear-mass measurements~\cite{McDonnell:2015sja}. We consider the Bayesian framework for neural networks formulated by MacKay \cite{MacKay_thesis}. It has been adapted to model electric and magnetic form  factors \cite{Graczyk:2010gw}. It was also used in the investigation of the two-photon exchange phenomenon in elastic electron-proton scattering~\cite{Graczyk:2011kh,Graczyk:2013pca,Graczyk:2014coa}. Furthermore, this approach has proved valuable  to gain insight into the proton radius puzzle and, in particular, to study the model dependence in the extraction of the proton radius from the electron-scattering data~\cite{Graczyk:2014lba,Graczyk:2015kka}.

In the present paper we employ the Bayesian framework for neural networks to find the most favorable $F_A (Q^2)$ based on the neutrino-deuteron CCQE scattering data measured by the ANL experiment\footnote{A global analysis including data from other experiments will be addressed in a subsequent study.}. In Sec.~\ref{Sec:Bayesian} the Bayesian approach for the feedforward neural networks is introduced. Section \ref{Sec:ANL_and_theo} introduces the ANL data and theoretical framework which describes the neutrino-deuteron scattering. In Sec.~\ref{Sec:Results} the numerical results are presented and discussed. In Sec.~\ref {Sec:Summary} the summary of the study is given. Appendix \ref{Appeendix_Prior} contains some details about the prior distributions in the Bayesian approach, whereas Appendix \ref{App:Fits} contains the analytic form of some of the fits.

\section{Bayesian framework for neural networks}

\label{Sec:Bayesian}

This section reviews the Bayesian approach formulated for the feedforward neural networks  and its adaptation to the problem of the extraction of the nucleon axial form factor that best represents a given set of data. The proposed framework is quite general: It does not rely on physics assumptions about the functional form of $F_A(Q^2)$ and is independent of the experimental conditions from which the data originate. In this way, the present approach not only complements those based on physically motivated parametrizations, but also has the potential to disclose new physics effects as well as deficiencies in the theoretical modeling or in the data. 

The general idea is the following:  Given a data set a statistical model is built. The model is characterized by a number of probability densities, which are obtained using feed-forward neural networks. A detailed account of the different ingredients of the approach is given in this section. The specific application to ANL CCQE neutrino-deuteron scattering data is left for the subsequent sections.

\subsection{Neural networks}
\label{Sec:Bayesian:sub:Feed-Forward}

Our aim is to obtain a statistical model which has the ability to generate $F_A(Q^2)$ values together with uncertainties.  In practice, to construct such a  model, a number of probability densities must be estimated. This can be achieved within  three general methods~\cite{Bishop_book}: (i) non-parametric, (ii) parametric, and (iii) semiparametric. In the first approach, no particular functional model is assumed, and the probabilities are determined only by the data. However, if the size of the data is large, the method requires introduction of many internal parameters. Additionally, this approach is computationally expensive.  In the parametric method, a specific functional form of the model is assumed. In  this case, it is relatively easy to find the optimal configuration of the model parameters. However, a particular choice of the parametrization limits the ability of the model for an accurate description of the data\footnote{Fitting the axial form factor with the dipole parametrization is an example of the parametric approach.}. In this case, the uncertainties for the model prediction are either overestimated or underestimated. The semiparametric method takes the best features from both (i) and (ii) approaches.  In this method, instead of a single specific functional model, a broad class of functions is considered. The optimal model is chosen among them. The neural-network approach is a realization of the semiparametric method. In particular, the feedforward neutral networks form a class of functions with unlimited adaptive abilities.

\subsection{Multilayer perceptron}

\label{Sec:Bayesian:sub:MLP}

\begin{figure}[htb]
\begin{center}
\includegraphics[width=0.5\textwidth]{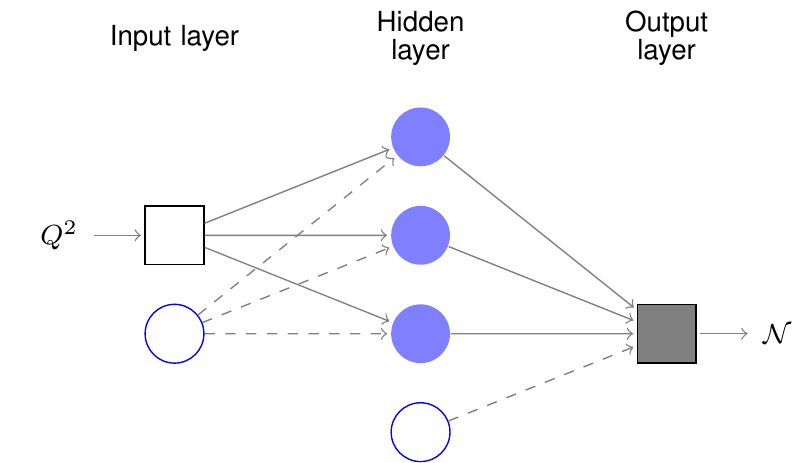}
\end{center}
\caption{
Feedforward neural network in an MLP configuration, $\mathcal{N}$:~$\mathbb{R} \mapsto \mathbb{R}$. It consists of: (i) an input layer with one input unit $Q^2$ (open square); (ii) one hidden layer with three hidden $M=3$ units (filled blue circles); (iii) an output layer  consisting of one output unit (black square). Each line denotes a weight parameter $w_j$. The bias weights are denoted by dashed lines, whereas the bias units  are represented by open blue circles.
\label{Fig_net}}
\end{figure}

In order to model the nucleon axial form factor a feed-forward neural network in a multilayer perceptron (MLP) configuration is considered. The concept of MLP comes from neuroscience~\cite{Rosenblatt62}. A given MLP is a nonlinear map from the input space of dimension $n_i$ to the output space of dimension $n_o$,
\begin{equation}
\mathcal{N}: \mathbb{R}^{n_i} \mapsto \mathbb{R}^{n_o}.
\end{equation}
The MLP map can be represented by a graph which consists of several layers of units: the input layer with $n_i$ units, one or more hidden layers with hidden units, and the output layer which has $n_o$ units. In the input and in every hidden layer, there is an additional bias unit. The units from the consecutive layers are all connected, but the bias unit is connected only to the following layer. As an example, the graphical representation of the MLP: $\mathcal{N}$: $\mathbb{R} \mapsto \mathbb{R}$ is given in Fig.~\ref{Fig_net}. Every edge (connection line) in the graph represents one parameter of the function, called latter a weight. 

\begin{figure}[htb]
	\begin{center}
	\includegraphics[scale=1]{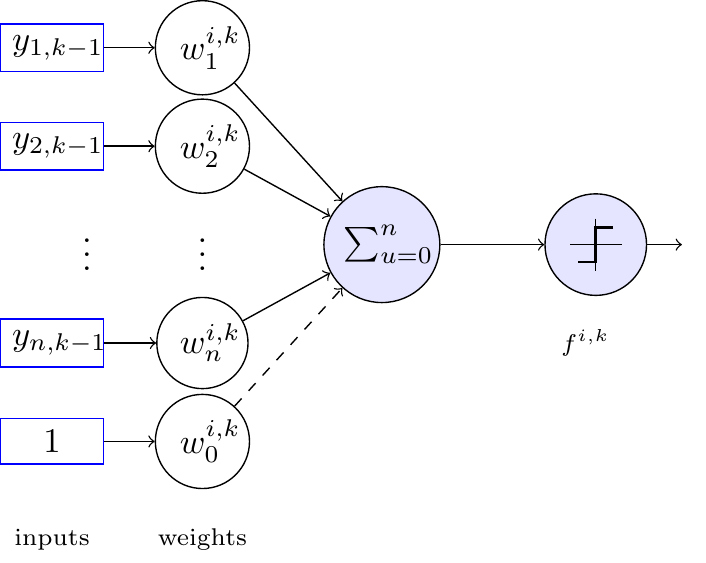}
	\end{center}
	\caption{The $i$th unit in the $k$th layer [Eq.~(\ref{Eq:unit})].   \label{Fig_unit}}
\end{figure}

To every unit (blue circles in Fig.~\ref{Fig_net}), a real single-valued function called the activation function $f$ is associated; its argument is the weighted sum of the activation function values received  from the connected units. In the feedforward case,  the $i$th unit in the $k$th layer  is given in terms of the input from the $(k-1)$th layer by  
\begin{equation}
\label{Eq:unit}
y_{i,k} = f^{i,k} \left(\sum_{u \in \mathrm{previous}\;\mathrm{layer}} w^{i,k}_u y_{u,k-1} \right).
\end{equation}
A graphical representation of the above function is given in Fig. \ref{Fig_unit}. The weights $w^{i,k}_u$ are real parameters. Their optimal values are established by the network training for which we adopt the Bayesian framework explained below. 

Note that for the bias unit $f^{i,k}(x) =1$. Furthermore, it is assumed that in the output layer the activation functions are linear $f(x) =x$.  In order to simplify and speed up the performance of the numerical analyses  MLPs with only one hidden layer of units are considered. In Fig.~\ref{Fig_net}, there is an example of such an MLP with $M=3$ hidden units.

Let us introduce the MLP $\mathcal{N}_M : \mathbb{R} \mapsto \mathbb{R}$, with a single hidden layer and $M$ units, which has the following functional form:
\begin{equation}
\label{Eq:functional:form:of:MLP}
\mathcal{N}_M(Q^2;\{ w_j \}) = 
\sum_{n = 1}^{M}  w_{2M+n }\, f \left(  w_{n} \frac{Q^2}{Q_0^2} + w_{M+n}\right)  + w_{3 M + 1} \,,
\end{equation}
where $Q_0^2 \equiv 1$~GeV$^2$.
This function depends on $W=3 M +1$ weights and $Q^2$.

It has been proved and demonstrated (Cybenko theorem)~\cite{Cybenko_Theorem,Hornik89,FUNAHASHI1989183,118638,80265,ITO1991385,KREINOVICH1991381} that, 
if $M$ is sufficiently large, feedforward neural networks with sigmoidal and linear activation functions in the (single) hidden and output layers, respectively, form a dense subset in the set of continuous functions. This implies that a map of the form (\ref{Eq:functional:form:of:MLP}) can approximate arbitrarily well any continuous function and its derivative. As required by the theorem, in our numerical analysis, the activation functions in the hidden layer are given by sigmoids,
\begin{equation}
\label{Eq:sigmoid}
f(x) = \frac{1}{1 + \exp(-x)}.
\end{equation}

\subsection{Axial form factor}

\label{Sec_Axial_form_factor}

We seek for a model-independent parametrization of $F_A$ that best represents the available data without any input from theory. It should be quite general but, nonetheless, constrained by the following general properties: 
\begin{enumerate}[(I)]
\item \label{assumption2} $F_A(Q^2)$ is assumed to be a continuous function of $Q^2$ in its validity domain; 
\item \label{assumption1} the domain of $F_A$ is restricted to the $Q^2 \in (0,3)$~GeV$^2$ where the ANL data are present;
\item \label{assumption3} $F_A(Q^2=0)$ is constrained by the $g_A$ experimental value, Eq.~\ref{Eq:ga_with_error};

\item \label{assumption4} as $F_A(Q^2)$ is bounded, there must be a $C>1$: $F_A(Q^2) < C F_A^{\mathrm{dipole}} (Q^2)$ in the whole $Q^2$ interval of $(0,3)$~GeV$^2$.
\end{enumerate}

The feedforward neural network of Eq.~(\ref{Eq:functional:form:of:MLP}) can fulfill these properties;  for a more detailed discussion see Appendix \ref{Appeendix_Prior}. In order to speed up the numerical computations, we rescale the output of the MLP (\ref{Eq:functional:form:of:MLP}) by normalizing it to the dipole ansatz. As the result, the axial form factor is represented by  
\begin{equation}
\label{Eq:FA_NN}
F_A (Q^2) = F_A^{\mathrm{dipole}} (Q^2) \times \; \mathcal{N}_M(Q^2;\{w_i\})\,,
\end{equation}
where $F_A^{\mathrm{dipole}}$ is given in Eq.~(\ref{Eq:FA_dipole}) with $M_A=1$~GeV. In this way, the neural-network response gives the deviation of the axial form factor from the dipole parametrization. The value of $g_A$ is allowed to change within the PDG uncertainty Eq.~(\ref{Eq:ga_with_error}).

\subsection{Bayesian approach and Occam's razor}

\label{Sec:Bayesian:sub:Occam}

As described above, a MLP is a nonlinear map defined by some number of the adaptive parameters. The increase in the number of hidden units improves MLP's ability to reproduce the data. However, when the number of units (weights) is too large, the model tends to overfit the data, and it reproduces the statistical noise. As a result, its predictive power is lost. On the other hand, if the network is too small, then the data are underfitted. This competition between two extreme cases is know in statistics as the bias-variance trade-off~\cite{Geman.1992.4.1.1}. Certainly, the optimal model is a compromise between both extreme situations. 

Bayesian statistics provides methods to face the bias-variance trade-off dilemma.  Indeed, the Bayesian approach naturally embodies Occam's razor \cite{MacKay_thesis,DAgostini_book,Berger1992,Jefferys_bis_1992} i.e. complex models, defined by a large number of parameters, are naturally penalized, whereas simple fits with a small number of parameters, are  preferred. Moreover, the Bayesian approach allows one  to make comparisons between different statistical descriptions of the data and to indicate the model which is favored by the measurements. An example of such analysis can be found in Ref. \cite{Graczyk:2014lba} where a large number of different fits of the electric and magnetic form factors were obtained from electron-proton scattering data. For each  model, the value of the proton radius $r_p^E$ has been calculated. It turned out that, depending on the model,  $r_p^E$ ranges from $0.8$ to $1.0$~fm. A considerable fraction of the results agreed with the  muonic-atom measurement $r^E_p=0.84184(67)$~fm~\cite{Pohl:2010zza}, but the Bayesian algorithm preferred a model with $r_p^E=0.899\pm0.003$~fm, which is in contradiction with the muonic-atom result but in agreement with some other non-Bayesian $ep$ scattering data analysis. A critical review of various approaches to proton radius extraction can be found in Ref.~\cite{Sick:2018fzn} and references therein.

\subsection{Bayesian framework for MLP}
\label{Sec:Bayesian:sub:BayesianforMLP}

\subsubsection{General idea}

We adopt the Bayesian framework for the feedforward neural network formulated by MacKay \cite{MacKay1992.4.3.415,MacKay1992.4.3.448}.  The main concepts of the approach are briefly reviewed below. 

Let us consider the set of neural networks 
\begin{equation}
\mathcal{N}_1, \, \mathcal{N}_2, \dots,\, \mathcal{N}_M,
\end{equation}
where $M$ denotes the number of units in the hidden layer. To each of the models $\mathcal{N}$, one associates a prior probability denoted $\mathcal{P}(\mathcal{N})$. Our task is to obtain two posterior conditional probabilities: $\mathcal{P}(\mathcal{N}\mid \mathcal{D})$ -- the probability of the model $\mathcal{N}$ given a data set $\mathcal{D}$ and  $\mathcal{P}(\rho \mid \mathcal{D},\mathcal{N})$ -- probability distribution of the model parameters given $\mathcal{D}$  and model assumptions; $\rho$ denotes the set of model parameters, which encompasses the neural network weights $\rho = \{ \{ w_j\}, \dots\}$. 
The first probability density allows us to choose among many network types the one which is favorable by the data, whereas the second one is necessary to make model predictions.  

If one assumes, at the beginning of the analysis, that all MLP configurations are equally suited for describing the data, then the following relations between prior probabilities hold 
\begin{equation}
\mathcal{P}(\mathcal{N}_1)=\mathcal{P}(\mathcal{N}_2)=\dots=\mathcal{P}(\mathcal{N}_M).
\end{equation} 
Then, in order to classify the models, it is sufficient to compute  the so-called evidence  $\mathcal{P}(\mathcal{D}\mid \mathcal{N})$. 
Indeed, from the Bayes' theorem, one gets
\begin{equation}
\label{eq:PND}
\mathcal{P}(\mathcal{N}\mid \mathcal{D}) = \frac{\mathcal{P}(\mathcal{D}\mid \mathcal{N})\mathcal{P}(\mathcal{N})}{\mathcal{P}(\mathcal{D})} \sim \mathcal{P}(\mathcal{D}\mid \mathcal{N}),
\end{equation}
where $\mathcal{P}(\mathcal{D})$ is the normalization constant.

On the other hand, the posterior probability for the weights of a given MLP reads
\begin{equation}
\label{eq:PwD_posterior}
\mathcal{P}(\rho\mid \mathcal{D} , \mathcal{N}) = \frac{\mathcal{P}(\mathcal{D}\mid \rho, \mathcal{N}) \mathcal{P}( \rho\mid \mathcal{N}) }{\mathcal{P}(\mathcal{D}\mid \mathcal{N})},
\end{equation}
where $\mathcal{P}(\mathcal{D}\mid \rho, \mathcal{N})$ is the likelihood whereas the density $\mathcal{P}(\rho\mid\mathcal{N})$ is the prior  describing the initial assumptions about the  parameters. By integrating both sides of Eq.~(\ref{eq:PwD_posterior}), one gets the evidence for the model,
\begin{equation}
\label{eq:PDN}
\mathcal{P}(\mathcal{D}\mid \mathcal{N}) = \int d \rho \, \mathcal{P}(\mathcal{D}\mid \rho, \mathcal{N})
\mathcal{P}( \rho \mid \mathcal{N}).
\end{equation}

\subsubsection{Likelihood, prior, and posterior densities}

In order to calculate the posterior (\ref{eq:PwD_posterior}), we assume that the likelihood is given in terms of the $\chi^2$ function,
\begin{equation}
\label{Eq:likelihood}
\mathcal{P}(\mathcal{D}\mid \{w_j\}, \mathcal{N} )  = \frac{1}{N_L}\exp( - \chi^2),
\end{equation} 
where $N_L$ is the normalization constant. The $\chi^2$ function for the present paper is defined in Sec.~\ref{Sec:Bayesian:sub:chi2} [see Eq. \ref{eq:chi2}]. 

It is also assumed that the initial values of the weights are Gaussian distributed (the arguments supporting this choice are collected in Appendix \ref{Appeendix_Prior})
\begin{equation}
\label{Eq:Prior_Ew}
\mathcal{P}(\{w_j\},\mathcal{N}) = \left(\frac{\alpha}{2\pi}\right)^{W/2} \exp\left( - \alpha\, E_w(\{w_j\})\right),
\end{equation}
where $\alpha$ is a hyperparameter (regularizer) introduced to deal with the overfitting problem and
\begin{equation}
\label{Eq:Eq}
E_w(\{w_j\})  =  \frac{1}{2}\sum_{i=1}^W w_i^2.
\end{equation}
The regularizer $\alpha$ plays a crucial role in model optimization and should be properly determined. Indeed, if $\alpha$ is large then the term (\ref{Eq:Eq}) dominates in the posterior Eq.~(\ref{eq:PwD_posterior}), so it is very likely that the model  underfits the data. On the contrary, if $\alpha$ is too small, the likelihood dominates, and the model tends to overfit the data. Note that $\alpha$ is another parameter of the model, hence, $\rho =\{ \{w_j\},\alpha \}$. 

In principle, to get the evidence $\mathcal{P}(\mathcal{D}\mid \mathcal{N})$ on which model discrimination is based, the integration in Eq.~(\ref{eq:PDN}) over the whole space of parameters $\rho$ should be performed. This is, however, numerically difficult to perform. Therefore, in our analysis, we consider another method, the so-called evidence approximation~\cite{MacKay1992.4.3.415,MacKay1992.4.3.448}.

\subsubsection{Evidence approximation}

In the adopted approach it is assumed that the posterior distributions have a Gaussian shape. 
Hence, to get the necessary information about (\ref{eq:PwD_posterior}) it is enough to obtain  the configuration of the parameters $\rho_{MP}$ at which the posterior distribution is at its maximum and the covariance matrix for the model. The latter is necessary to provide the uncertainties for the model predictions.

In this approach, which corresponds to the type II maximum likelihood method of conventional statistics \cite{Berger_1985}, the optimal value  $\rho_{MP}$ is established during the training of the network. 
In this process, the \textit{a priori} unknown $\alpha$ parameter is iteratively changed \{see Eqs. (3.18) and (3.19) of Ref.~\cite{Graczyk:2010gw}\}, starting from small value $\alpha_0=0.001$, which leads to a posterior covering a large region in the parameter space. 
The iteration procedure is convergent, and the result has a negligible dependence on the initial $\alpha$ value. More details about the algorithm implementation can be found in Section 3.1 of Ref.~\cite{Graczyk:2010gw} and in Sec. III. C of \cite{Graczyk:2013pca}.
Note that the optimal configuration of the model parameters  $\rho_{MP} = \{\{w_j\}_{MP},\alpha_{MP}\}$ is close to the configuration for which  the  $\chi^2$ is at the minimum.

Within the present approximation, the evidence for a given model is cast in an analytical form. Namely, 
\begin{eqnarray}
\ln \mathcal{P}\left(\mathcal{D}\right|\left. \mathcal{N} \right)
\label{Eq:log_of_evidence_misfit}
 &\approx&   -\chi^2 - \alpha_{MP} E_{w}(\{w_j\}_{MP})  \\ [0.2cm]
&-&\frac{\ln \mid A \mid}{2}  + \frac{W}{2}\ln \alpha_{MP} -\frac{1}{2}\ln \frac{\gamma(\alpha_{MP})}{2}  \nonumber \\[0.2cm]
&+& M \ln(2) + \ln(M!)\,. 
\label{Eq:log_of_evidence}
\end{eqnarray}
In the above expression, normalization factors common to all models are omitted;  $|A|$ denotes the determinant of the Hessian  matrix, 
\begin{equation}
A_{ij} = \nabla_i \nabla_j \left. \chi^2 \right|_{\{w_k\}=\{w_k\}_{MP}} + \delta_{ij}\alpha_{MP}\,.
\end{equation} 
The parameter $\gamma$ is given by 
\begin{equation}
\gamma(\alpha) = \sum_{i=1}^W \frac{\lambda_i}{\alpha + \lambda_i};
\end{equation}
$\gamma(\alpha_{MP})$ measures the effective number of weights, whose values are controlled by the data \cite{Bishop_book}. The $\lambda_i$'s are eigenvalues of the matrix $\nabla_i\nabla_j \chi^2$.

\begin{figure}[H]
\begin{center}
	\includegraphics[width=0.49\textwidth]{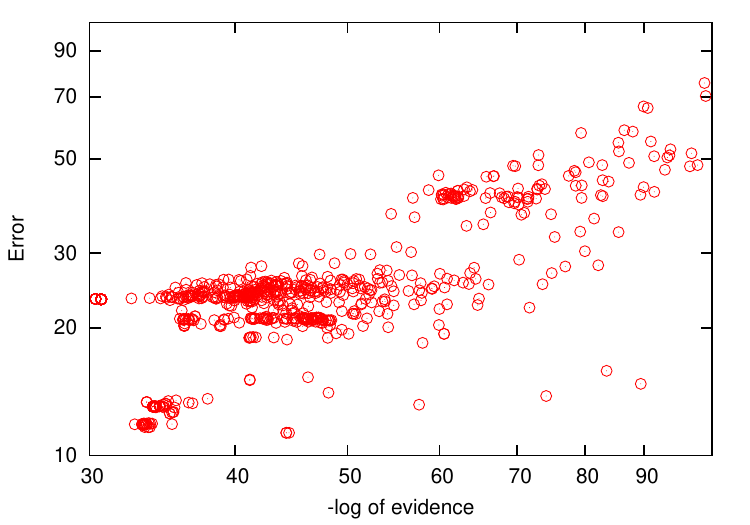}
\end{center}
\caption{The error Eq.~(\ref{Eq:Error_function}) as a function of the logarithm of the evidence Eq.~(\ref{Eq:log_of_evidence_misfit} and \ref{Eq:log_of_evidence}). Each point denotes the result obtained for one MLP fit to the BIN1 data, including deuteron corrections (see Sec.~\ref{Sec:ANL_and_theo} for details). The fits with a logarithm of the evidence smaller than $-100$ are not shown in the figure. 
\label{Fig_error_vs_evidence}}
\end{figure}

The evidence contains two contributions: Occam's factor [(\ref{Eq:log_of_evidence}) plus the $\alpha_{\text{MP}} E_w$ term of \cite{Hill:2017wgb}],  which is large for models with many parameters and the misfit [$\chi^2$ term in Eq.~(\ref{Eq:log_of_evidence_misfit})], which could be large if the model is too simple. Therefore, the model which maximizes the evidence is the one which solves the bias-variance dilemma. As an illustration from the present paper (details can be found in the following section),  in Fig.~\ref{Fig_error_vs_evidence} we plot the values of the error, 
\begin{equation}
\label{Eq:Error_function}
\mathcal{E}  =  
\chi^2
+ \alpha E_w ,
\end{equation}
and the evidence for MLP fits. The best model with the highest evidence is not the one which has the smallest value of the error function $\mathcal{E}$, in variance with more conventional approaches based on the minimization of the $\chi^2$ per degree of freedom.

\section{Analysis of ANL neutrino-deuteron scattering data}

\label{Sec:ANL_and_theo}

\subsection{Theoretical framework}

The neutrino-induced CCQE, 
\be
\label{eq:CCQEreaction}
\nu_\mu(k) + n(p) \raw \mu^-(k') + p(p') ,
\ee
differential cross section, in terms of the standard Mandelstam variables $s=(k+p)^2$, $u=(p-k')^2$, and $t=(k-k')^2=-Q^2$, can be cast as~\cite{LlewellynSmith:1971uhs}
\bea
\label{eq:CCQE}
&&\frac{d\sigma}{dQ^2} = \frac{G_F^2 m_N^2} {8 \pi E_\nu^2}\times \nonumber \\
 && \left[ A(Q^2) +  B(Q^2) \frac{(s-u)}{m_N^2}+ C(Q^2) \frac{(s-u)^2}{m_N^4} \right]\,,
\eea
where $A$, $B$, and $C$ are quadratic functions of the vector [$F^V_{1,2}(Q^2)$] and axial [$F_{A,P}(Q^2)$] form factors \{see Eqs.~(3.18) of Ref.~\cite{LlewellynSmith:1971uhs}\}; $G_F$ is the Fermi constant, $m_N$ is the nucleon mass, and $E_\nu$ is the neutrino energy in the laboratory frame. Isospin symmetry allows to relate $F^V_{1,2}$ to the corresponding electromagnetic proton and neutron form factors, which are extracted from electron-scattering data. In the present paper we have taken these electromagnetic form factors from Refs.~\cite{Galster:1971kv,Nieves:2004wx}. With this simple choice we disregard deviations from the dipole shape because the accuracy of the neutrino-deuteron data is insufficient to be sensitive to them, particularly at the rather low $Q^2 \lesssim 1$~GeV$^2$ probed in the ANL experiment. Finally,  the partial conservation of the axial current (PCAC) and the pion-pole dominance of the pseudoscalar form factor $F_P$ allow to express it in terms of $F_A$.

Deuterium-filled bubble-chamber experiments actually measured $\nu_\mu + d \to \mu^- + p + p$. The cross section for this process differs from the one on free neutrons due to the momentum distribution of the neutron in the nucleus, Pauli principle, final-state interactions, and meson-exchange currents. In the literature, it has been commonly assumed that Eq.~(\ref{eq:CCQE}) can be corrected for these effects by a multiplicative function of $Q^2$ alone $R(Q^2)$ and such that $R \raw 1$ at large $Q^2$. For the present paper, we adopt $R(Q^2)$ from the calculation of Ref.~\cite{Singh:1986xh} (solid line in Fig.~4).

\subsection{$\chi^2$ function for the ANL experiment}

\label{Sec:Bayesian:sub:chi2}

In the ANL experiment, the interactions of muon neutrinos in a 12-ft bubble chamber filled with liquid deuterium were studied~\cite{Mann:1973pr,Barish:1977qk,Barish:1978pj,Miller:1982qi}. The neutrino flux peaked at $E_\nu \sim 0.5$~GeV and has fallen by an order of magnitude at $E_\nu =2$~GeV~\cite{Barish:1977qk,Miller:1982qi}. 
For the statistical analysis, we consider the $Q^2$ distribution of CCQE events. Some of the originally published bins were combined together to have a number of events larger than five in every bin. Therefore, the number of bins is $n_\mathrm{ANL}=25$, where bins from $1$ to $23$ have a width of $0.05$~GeV$^2$, whereas bins $24$ and $25$ have widths of $0.65$ GeV$^2$. The total number of measured two- and three-prong events adds to $N_\mathrm{ANL}= 1792$~\cite{Miller:1982qi}. One-prong events were not included in the ANL selection. To account for their loss, the region of $Q^2=0.05$~GeV$^2$ was excluded. 

The predicted number of events in each bin is calculated similarly as in Ref.~\cite{Meyer:2016oeg}, 
\begin{equation}
N_i^{th}=p \int^\infty_0 dE_\nu \frac{\displaystyle \frac{d\sigma}{d Q^2}(E_\nu,F_A,Q_i^2)}{\sigma (E_\nu,F_A)}\frac{dN}{dE_\nu}\,,
\label{Eq:N_th}
\end{equation} 
where $p (dN/dE_\nu)/\sigma (E_\nu,F_A) $ is the neutrino energy flux, given in terms of the experimental energy distribution of observed events $dN/dE_\nu$ taken from Ref.~\cite{Barish:1978pj}.

As stated in the previous section, the likelihood [Eq.~(\ref{Eq:likelihood})] is built using the $\chi^2$ function, which we cast as 
\begin{equation}
\label{eq:chi2}
\chi^2 =\chi^2_\mathrm{ANL} + \chi^2_{g_A},
\end{equation} 
where $\chi_{g_A}^2$ is introduced to constrain the value of the axial form factor at $Q^2=0$,
\begin{equation}
\label{Eq:chi2_gA}
\chi^2_{g_A} = \left( \frac{F_A(0) - g_A}{\Delta g_A}\right)^2\,;
\end{equation}
$g_A$ and $\Delta g_A$ are fixed by the present PDG central value and its uncertainty, respectively, Eq.~(\ref{Eq:ga_with_error}). For $\chi_\mathrm{ANL}^2$, we take  
\begin{equation}
\label{Eq:chi2_ANL}
\chi_\mathrm{ANL}^2= \sum_{i=k}^{n_\mathrm{ANL}} \frac{\left(N_i - N_i^{th}\right)^2}{N_i} + \left(\frac{1 - p}{\Delta p}\right)^2 \,,
\end{equation}			
where $N_i$ denotes the number of events in the bin. The last term takes into account the systematic uncertainty in the total number of events~\cite{DAgostini:1993arp} inherited from the flux-normalization uncertainty. Similarly as in the analysis of single pion production data~\cite{Graczyk:2009qm}, it is assumed that $\Delta p =0.20$~\footnote{This is a more conservative value of the flux-normalization uncertainty than the ANL estimate of 15\%~\cite{Miller:1982qi}.}. At the beginning of the analysis, $p=1$ is set. Then, during the training of the network, $p$ is iteratively updated. This algorithm is described in Ref.~\cite{Graczyk:2011kh}. 

It is known that the low-$Q^2$ data are characterized by a lower efficiency (see, for instance, Fig.~1 of Ref.~\cite{Miller:1982qi}). Moreover, in this kinematic domain deuteron structure corrections must be carefully discussed. In order to study this problem we consider three variants of the ANL data:
\begin{enumerate}[(i)]
	\item $\chi^2_\mathrm{ANL}\to\chi^2_{\mathrm{BIN}0}$: All ANL bins included;
	\item $\chi^2_\mathrm{ANL}\to\chi^2_{\mathrm{BIN}k}$: where $k=1$ or $k=2$: ANL bins without the first $k$ bins.
\end{enumerate}
Additionally,  for each data set, we consider the cross-section model both with and without [$R(Q^2)\equiv 1$] deuteron corrections.

\subsection{Numerical algorithm}
\label{Sec:Bayesian:sub:Algorithm}

We consider a MLP with $M=1-4$ hidden units in a single hidden layer. For $M>4$, the number of parameters in the fit starts to be comparable with number of bins. The numerical algorithm for getting the optimal fit  is summarized by the following list of steps: 
\begin{enumerate}[(i)]

\item Consider a MLP with a fixed number of hidden units $M=1$; 

\item using the Bayesian learning algorithm (\cite{Graczyk:2013pca}), perform the network training and find the optimal values for the weights and the regularizer $\alpha$;
\begin{itemize}
\item set the initial value of $\alpha\equiv\alpha_0=0.001$;
\item initialize randomly the values of the weights;
\item perform training until the maximum of the posterior is reached; at each iteration step update the values of weights and $\alpha$.
\end{itemize}

\item Calculate the evidence for each of the obtained MLP fits;

\item repeat steps (i) (iii) for various initial configurations of $\{w_j\}$; 

\item among all registered fits choose the best one according to the evidence;

\item repeat steps (i) (iv) for $M=2-4$;

\item among the best fits, obtained for $\mathcal{N}_{1-4}$ MLPs, choose the model with the highest evidence.
\end{enumerate}
The optimal configuration of parameters is obtained using the Levenberg-Marquardt algorithm \cite{Levenberg44,Marquardt63}.

\section{Numerical results}

\label{Sec:Results}

The analysis of the BIN0, BIN1, and BIN2 data sets has been independently performed. For each set, both cross-section models with and without deuteron corrections have been studied. For the default analyses, $\Delta g_A$ has been taken from the PDG as in Eq.~\ref{Eq:ga_with_error}, but the impact of a larger uncertainty $\Delta g_A/g_A = 10$\%  has been investigated and is discussed below. We have also performed analyses with normalization uncertainties smaller ($\Delta p= 0.10$) and larger ($\Delta p= 0.30$) than the default $\Delta p= 0.20$, but it turned out that decreasing or increasing $\Delta p$ does not significantly affect the final results. All in all, about 17000 fits have been collected. Among them, for each type of analysis, the best model has been chosen according to the objective Bayesian criterion described in Sec.~\ref{Sec:Bayesian}.

In order to compare quantitatively different analyses, one needs to take into account the relative data normalization $\mathcal{P}(\mathcal{D})$. This density is not evaluated within the adopted approach. Hence, we can not quantitatively compare the results of, e.g., BIN0 and BIN1 analyses. Nonetheless, for a given data set, quantitative comparisons between the results obtained within with the two versions of the cross-section model can be made.

As described in Sec. \ref{Sec:Bayesian:sub:Algorithm} for each type of analysis (data set plus cross section model), to find the optimal fit, MLPs with: $M=1,2,3$ and $4$, hidden units have been  trained. The best model within each MLP type is the one with the maximal value of the evidence, Eqs.~(\ref{Eq:log_of_evidence_misfit} and \ref{Eq:log_of_evidence}). 

\begin{figure}[htb]
	\begin{center}
		\includegraphics[width=0.5\textwidth]{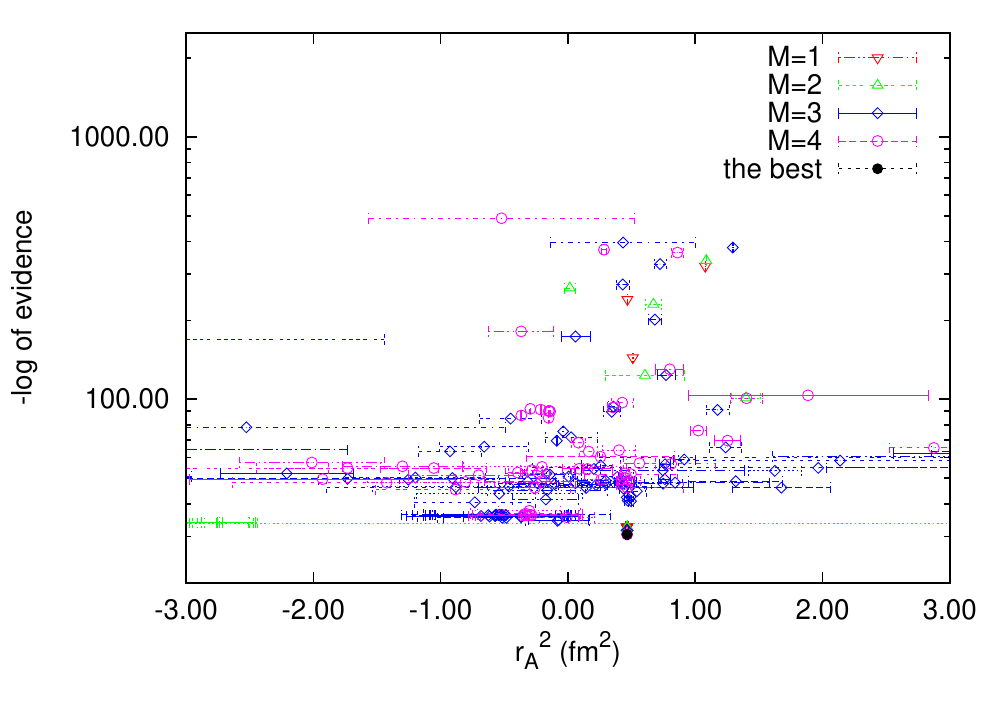}
	\end{center}
	\caption{The dependence of  $r_A^2$, defined in Eq.~(\ref{Eq:ra2}), on the logarithm of the evidence (\ref{Eq:log_of_evidence_misfit} and \ref{Eq:log_of_evidence}). Results for the MLP fits to BIN$1$ data (without the deuteron correction).  The MLPs consist of $M=1-4$ hidden units. \label{Fig_BIN1_all_rA2_fits}}
\end{figure}

In order to illustrate the performance of the training algorithm, in Fig.~\ref{Fig_BIN1_all_rA2_fits} we present the dependence of the resulting axial-radius squared ($r_A^2$) values on the evidence for the BIN1 data set.
The best fit with the highest evidence, obtained with $M=4$, gives $r_A^2 \approx 0.464$~fm$^2$.  

\begin{figure}[htb]
	\begin{center}
		\includegraphics[width=0.5\textwidth]{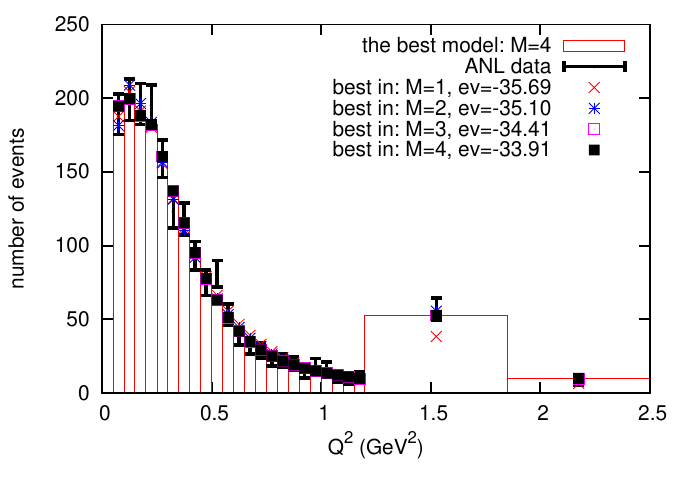}
	\end{center}
	\caption{Distribution of the ANL number of events and the best fits obtained for MLPs with $M=1-4$ hidden units. The figure shows the results of the analysis of BIN0 data (deuteron corrections included). For each fit, the value of the logarith of evidence (ev) is given.	
 \label{Fig_ANL_distribution_example}}
\end{figure}

Note that all the best models within each MLP type reproduce well the ANL data. This is illustrated in Fig.~\ref{Fig_ANL_distribution_example}, which presents the distribution  of the ANL events and the best fits. 

\begin{table}[htb]
	\begin{center}
		\begin{tabular}{|c|c|c|c|c|c|c|}
			\hline 
			Deuteron&  M & $\ln \mathcal{P}(\mathcal{D},\mathcal{N})$ & $\chi^2$ &  $p$  & $\mathcal{E}$ &  $r_A^2$ (fm$^2$) \\
			\hline
			\multicolumn{7}{|c|}{BIN0}\\ 
			\hline
			No & $4$ & $-34.72$ & $11.97$ & $1.1$ & $14.14$ & $-0.394\pm 0.278$\\
			Yes & $4$ & $-33.91$ & $11.73$ & $1.08$ & $13.95$ & $-0.161\pm 0.240$\\
			\hline
			\multicolumn{7}{|c|}{BIN1}\\
			\hline 
			No & $4$ & $-30.57$ & $24.84$ & $1.16$ & $25.41$ & $0.464\pm 0.014$\\
			Yes & $3$ & $-29.6$ & $22.90$ & $1.12$ & $23.43$ & $0.471\pm 0.015$\\
			\hline
			\multicolumn{7}{|c|}{BIN2}\\
			\hline 
			No & $2$ & $-30.15$ & $22.62$ & $1.18$ & $23.16$ & $0.476\pm 0.017$\\
			Yes & $4$ & $-27.67$ & $21.94$ & $1.13$ & $22.62$ & $0.478\pm 0.017$\\
			\hline
		\end{tabular}
		\caption{The best MLP fits, obtained for the analysis of the BIN0, BIN1 and BIN2 data with and without deuteron corrections; $\Delta g_A$ is taken from Eq.~\ref{Eq:ga_with_error}.   
			\label{Table_BINSPDG}}
	\end{center}
\end{table}

\begin{figure}[htb]
\centering{
\includegraphics[width=0.5\textwidth]{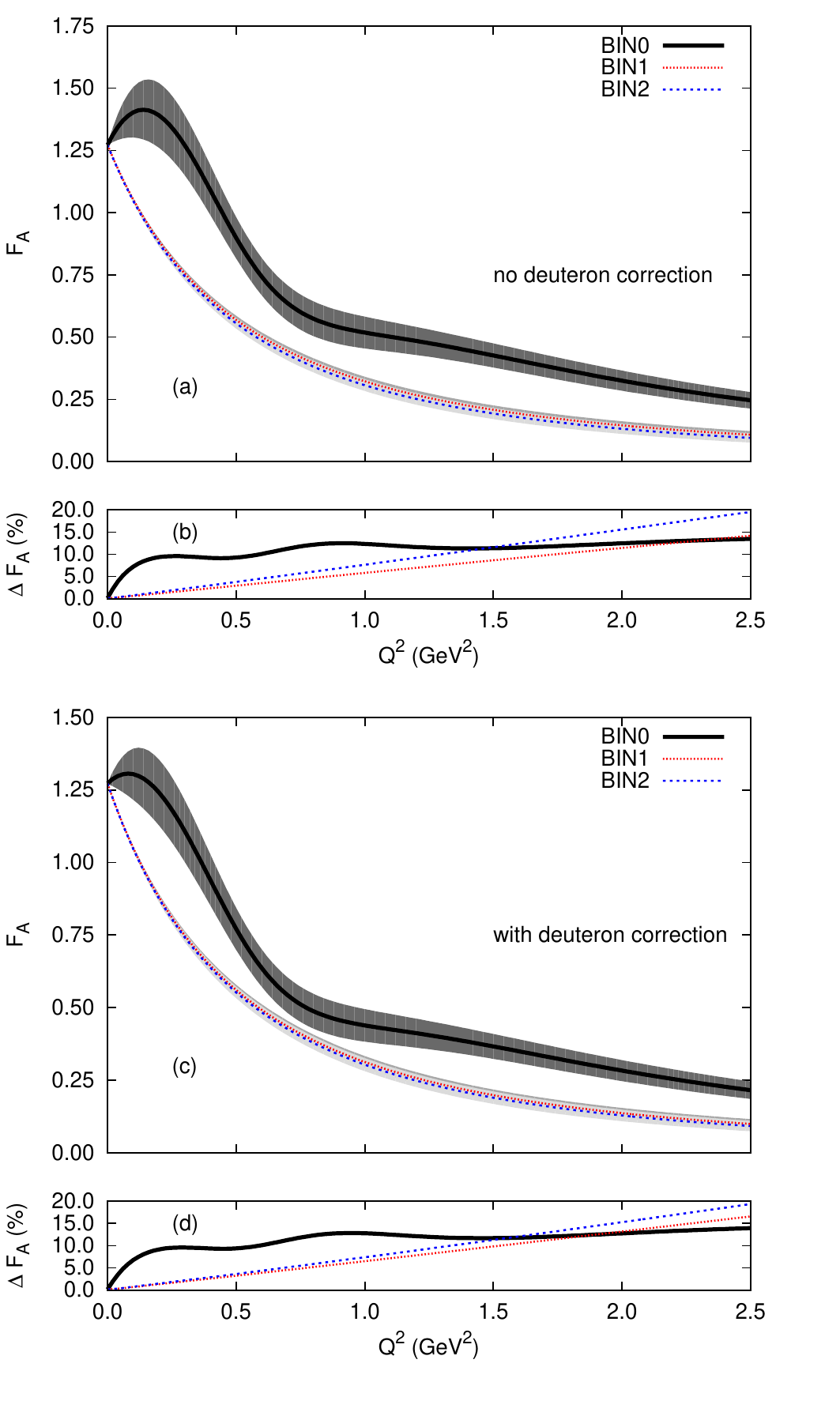}
}
\caption{Best fits of the axial form factor obtained from the analysis of the BIN0, BIN1 and BIN2 data sets. The top (bottom) panel presents the results  obtained without (with) deuteron corrections. The shaded areas denote $1\sigma$ uncertainties. Additionally, the relative uncertainty $\Delta F_A/F_A$ is plotted. 
\label{Fig_PDG_BINS_comparisons}}
\end{figure}

Our main results, i.e., the best fits to  BIN0, BIN1, and BIN2 data for the model with and without the deuteron correction with $\Delta g_A$ from Eq.~(\ref{Eq:ga_with_error}) are summarized in Table~\ref{Table_BINSPDG}. The corresponding $F_A(Q^2)$ functions and their error bands are shown in Fig.~\ref{Fig_PDG_BINS_comparisons}.    

Both fits for the BIN0 data set, which contains all the data from the original ANL measurement, with and without deuteron corrections show a $Q^2$ behavior of $F_A$ with a rapid increase followed by a decrease after a local maximum. As a result, $r_A^2$ has a negative sign\footnote{Although the large uncertainty does not exclude positive values.}, which is at odds with all available determinations. We also observe that the $Q^2$ dependence of the form factor disagrees with the one obtained from the same data set using the {\it z} expansion \{the coefficients from the ANL fit are given in Eq.~(19) of Ref.~\cite{Meyer:2016oeg}\}. We have obtained the {\it z} expansion coefficients for the BIN0 best fit, finding that their values grow with the expansion order to values that are too large compared to phenomenological expectations~\cite{Bhattacharya:2011ah}. This is an indication that the fit that best represents BIN0 data is inconsistent with the QCD assumptions implicit in the {\it z}expansion.

The height of the $F_A$ maximum is reduced once the deuteron correction is included in the analysis, and it disappears when the first bin is removed from the ANL data (BIN1 data set)\footnote{It is worth mentioning that fits with a negative slope of $F_A$ at low $Q^2$, resembling the best result for BIN0 data, have also been obtained in this case, but they are not preferred by the Bayesian algorithm.}. Hence, the presence of the local maximum of $F_A$ appears to be caused by low-$Q^2$ effects. Furthermore, the coefficients of the {\it z} expansion for the fits to BIN1 and BIN2 data sets are fully consistent with the expectations from QCD.

There are  several possible sources of this unexpected behavior of the fits  to the BIN0 set, namely, (i) an improper description of the nuclear corrections; (ii) a low quality of the measurements at low-$Q^2$ due to low and not well understood efficiency; (iii) constraints coming from the uncertainty of $g_A$; (iv) because of the lack of very low-$Q^2$ data, the actual value of $r_A^2$ might not be properly estimated: For instance, if $F_A$ has first a local minimum and then a local maximum \footnote{The magnetic form factors of the nucleon at  very low-$Q^2$ (about $ 0.01$~GeV$^2$) when normalized to a dipole have an oscillatory $Q^2$ dependence\cite{Graczyk:2010gw}.}. In the later scenario, the ANL data (and the available bubble-chamber data, in general) are not precise enough to reveal this behavior.

 \begin{figure}[htb]
 	\centering{
 		\includegraphics[width=0.5\textwidth]{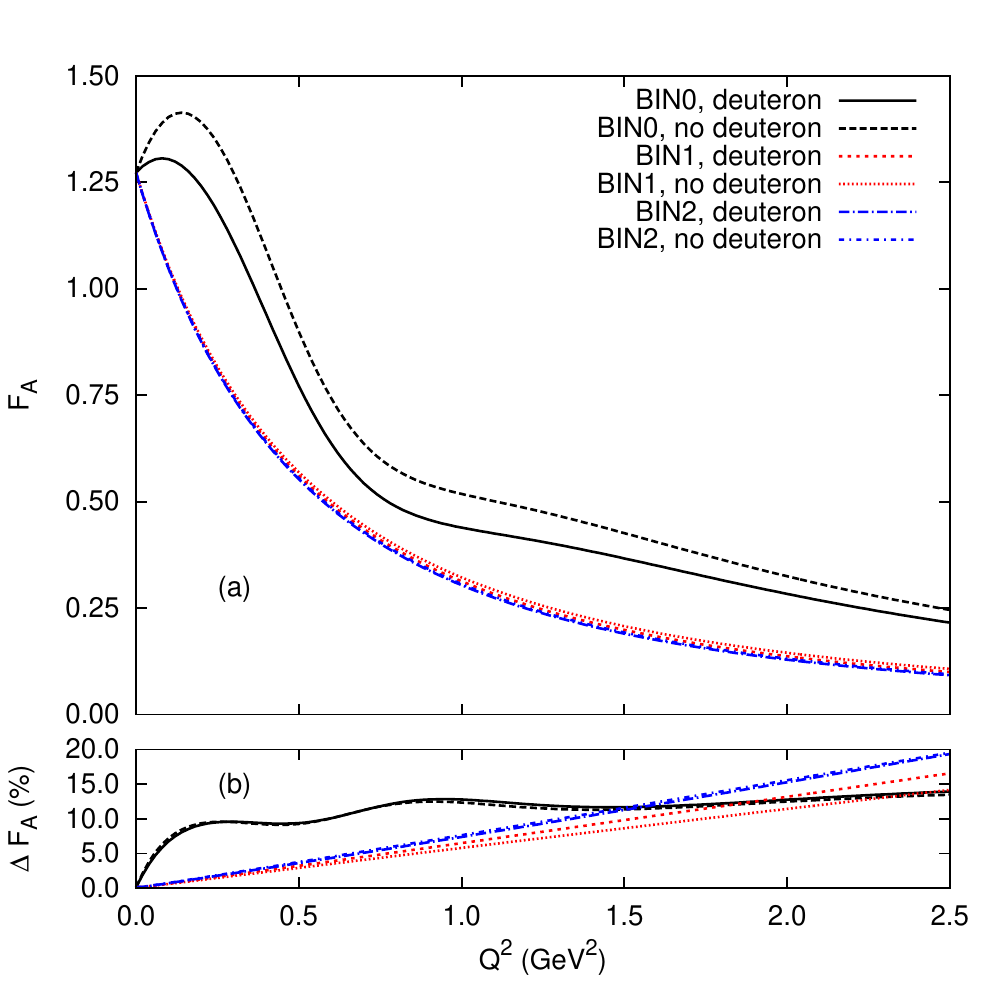}
 	}
 	\caption{Impact of the deuteron corrections on the axial form factor fits. Results of the best fits to the BIN0, BIN1, and BIN2 data sets with and without the deuteron correction together with relative uncertainties. All curves for the BIN1 and BIN2 cases nearly overlap.    
 \label{Fig_PDG_Singh_comparisons}
 	}
 \end{figure}
In the low-$Q^2$ kinematic domain, deuteron effects are sizable and may play a crucial role. On the other hand the inclusion of deuteron corrections in the analysis of the BIN1 and BIN2 data sets has a minor impact on the functional dependence of the final results, i.e., there is small difference between $F_A(Q^2)$ obtained with and without deuteron corrections as can be seen in Fig.~\ref{Fig_PDG_Singh_comparisons}. It is also interesting to highlight that the inclusion of the deuteron-structure corrections in the cross-section model increases the value of the evidence for each type of the analysis, see Table \ref{Table_BINSPDG}. 
 
  \begin{figure}[htb]
 	\begin{center} 		\includegraphics[width=0.5\textwidth]{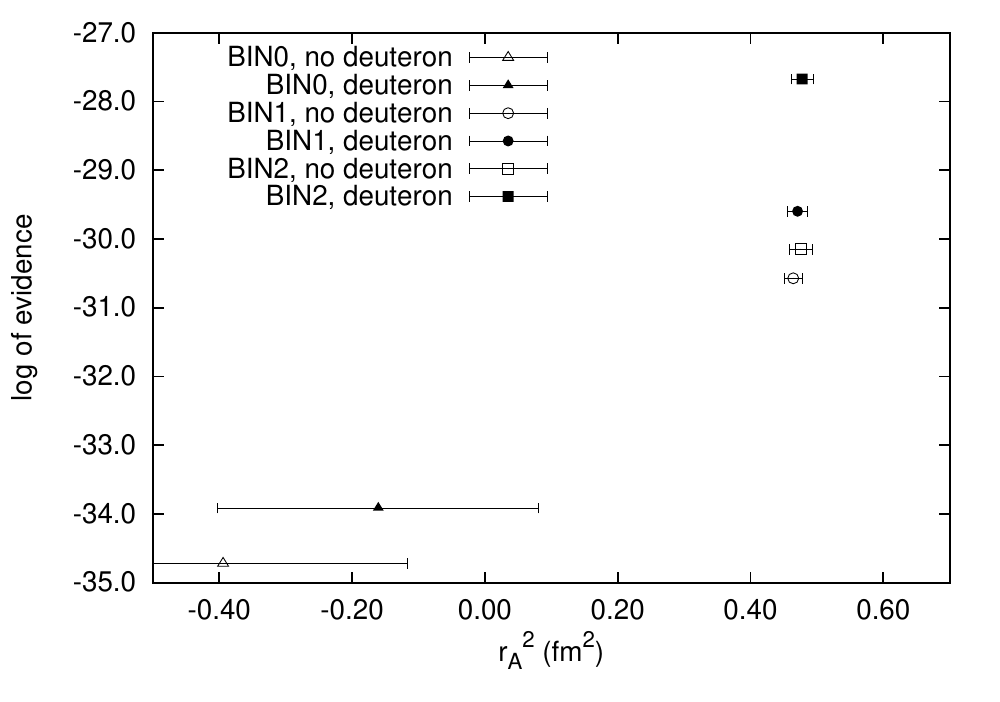}
 	\end{center}
 	\caption{Dependence of $r_A^2$ on the evidence. Open and full triangles denote (the best) fits to the BIN0 data without and with the deuteron corrections, respectively. Analogously, the fits to the BIN1 and BIN2 data are denoted by circles and squares, respectively.
\label{Fig_ra2_evidence}}
 \end{figure} 
 
 In  Fig.~\ref{Fig_ra2_evidence}, we plot values of $r_A^2$ against the evidence. It is clearly seen that the fits including deuteron corrections are favored by the ANL data. The impact of the sensitivity of the results on the deuteron structure revealed in the present paper calls for a more accurate account of this ingredient of the cross section models, beyond the $R(Q^2)$ function from Ref.~\cite{Singh:1986xh} employed so far. Recent studies of CC $\nu d$ scattering in the QE regime (without pions in the final state) include the nonrelativistic calculation of the inclusive cross section, incorporating two-body amplitudes, of Ref.~\cite{Shen:2012xz}. For the kinematics of the ANL and other bubble-chamber experiments, it is important to employ a relativistic framework as in Ref.~\cite{Moreno:2015nsa}. Furthermore, the consideration of the semi-inclusive rather than the inclusive cross section will allow taking into account the detection threshold for outgoing protons, which, in the ANL case, is 100~MeV~\cite{Barish:1977qk}. One should nonetheless bear in mind that even with the best model for the deuteron there is no guarantee that the low-$Q^2$ region is successfully described because of the difficulties in the measurement and with efficiency estimates at this kinematics.       

\begin{figure}[htb]
\bcen
 		\includegraphics[width=0.5\textwidth]{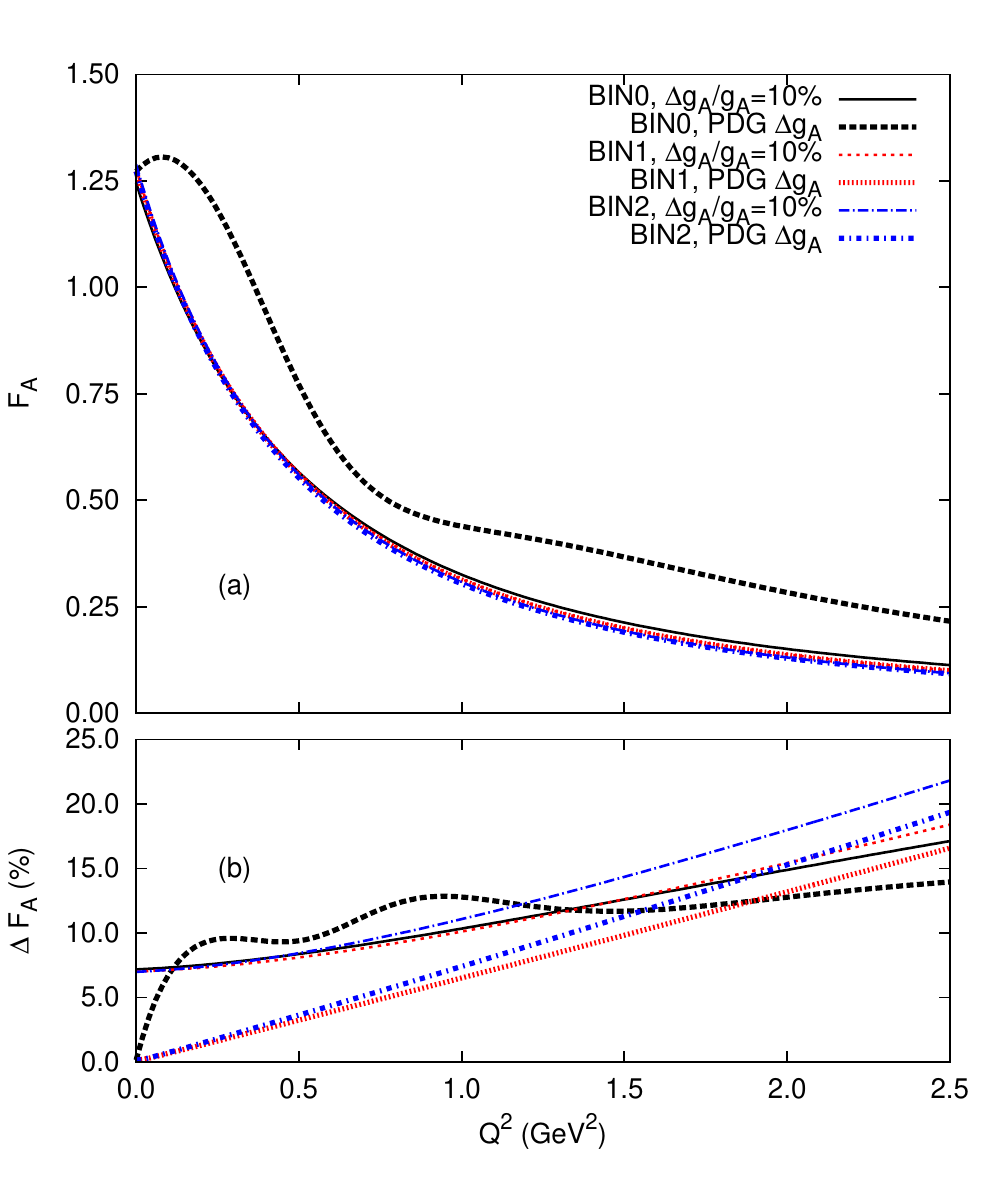}
\ecen
 	\caption{Impact of the $\Delta g_A$ uncertainty on the extraction of the axial form factor. The best fits to  BIN0, BIN1 and BIN2 data (with deuteron correction).  The thin lines denote results obtained assuming a $10\%$ uncertainty for $g_A$, whereas the thick lines denote fits with the PDG uncertainty.
 \label{Fig_10vsPDG_Singh_comparisons}}
 \end{figure}

 The impact of $\Delta g_A$ on the results can be easily investigated. Indeed  if one increases the $\Delta g_A$ uncertainty from $\Delta g_A/g_A \approx 0.1$\%, as in Eq.~(\ref{Eq:ga_with_error}) to  $10$\%, then the local maximum of $F_A$ disappears.  However, the fit uncertainty rapidly grows from $\Delta F_A/F_A$ lower than $0.01\%$  to $\Delta F_A/F_A \approx 7\%$ at $Q^2=0$. This analysis is shown in Fig.~\ref{Fig_10vsPDG_Singh_comparisons}.

\begin{figure}[h]
  	\begin{center}
  		\includegraphics[width=0.5\textwidth]{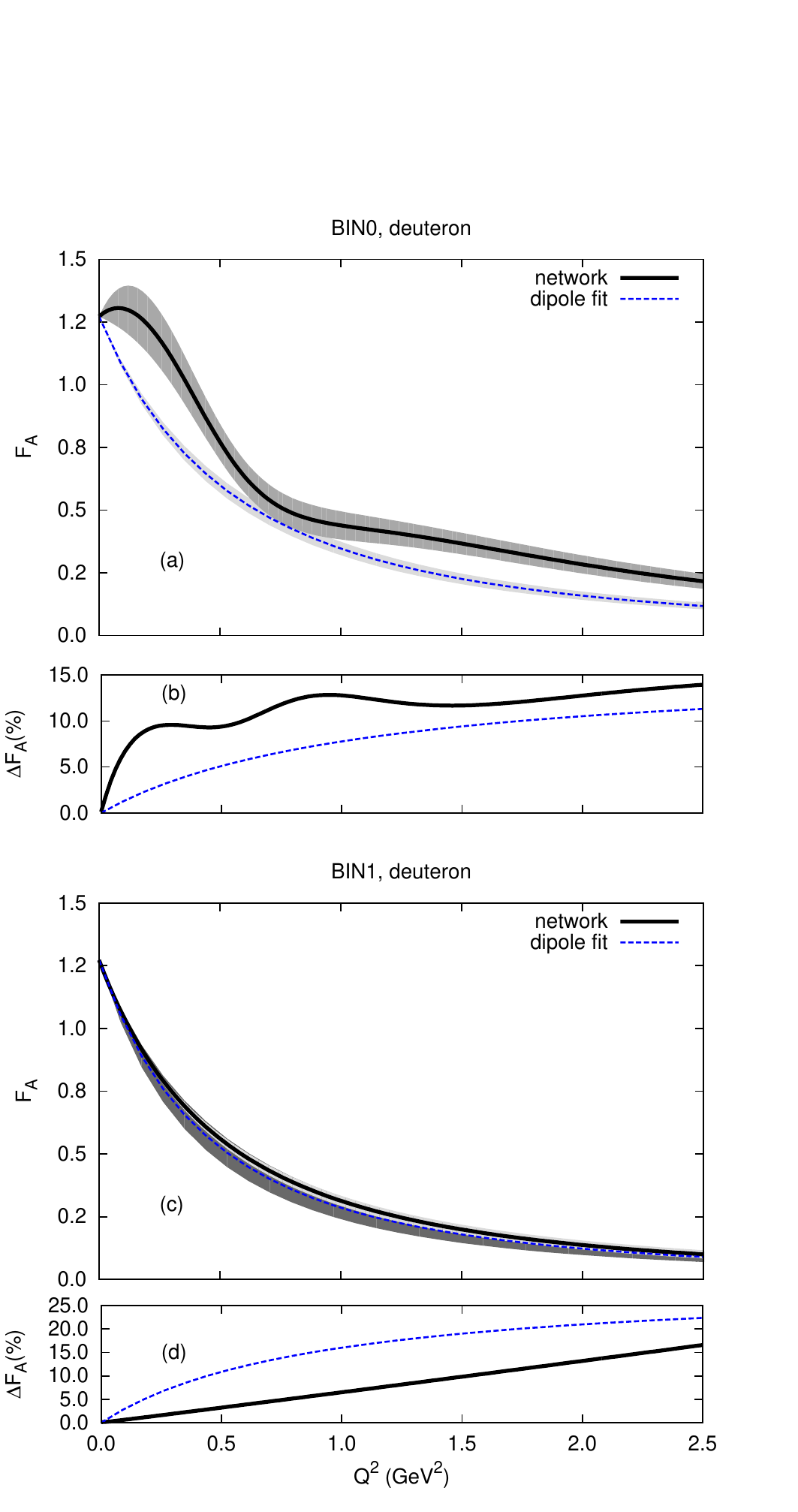}
  	\end{center}
  	\caption{Comparison of the dipole with the neural network fits to the BIN0 and BIN1 data, the deuteron corrections included. The shaded areas denote $1\sigma$ uncertainties of $F_A$.
  	\label{Fig_BIN0_dipole_vs_net}}
  \end{figure}

In order to compare the Bayesian neural-network results with the traditional approach, we have performed a conventional analysis of the ANL data assuming the dipole parametrization for the axial form factor Eq.~\ref{Eq:FA_dipole}. The best fit minimizes the  $\chi^2_{\mathrm{ANL}}$ function [Eq.~\ref{Eq:chi2_ANL}]\footnote{For these analyses, the MINUIT package of ROOT has been utilized.}. These results are summarized in Table~\ref{Table_Minuit_ANL}, whereas the comparison between dipole fits and neural-network analyses are displayed in Fig.~\ref{Fig_BIN0_dipole_vs_net}.

\begin{table}[htb]
	\begin{center}
		\scalebox{0.9}{
			\begin{tabular}{| l | c | c | c | c |}			
				\hline
				\multicolumn{5}{|c|}{$\chi^2$ analyses}\\ \hline
				& $\chi^2$ & $p$ & $M_A \:(\mathrm{MeV})$  & $r^2_A \:(\mathrm{fm}^2)$\\  
				\hline                                        \multicolumn{5}{|c|}{}\\ 
				\multicolumn{5}{|c|}{BIN0}\\ \hline
				No deuteron  & 33.3 & 1.12 $\pm$ 0.03 & 1110 $\pm$ 60  & 0.38 $\pm$ 0.04\\ \hline
				Deuteron  & 28.0 & 1.09 $\pm$ 0.03 & 1050 $\pm$ 60  & 0.43 $\pm$ 0.05\\  
				\hline
				\multicolumn{5}{|c|}{}\\
				\multicolumn{5}{|c|}{BIN1}\\
				\hline
				& $\chi^2$ & $p$ & $M_A \:(\mathrm{MeV})$  & $r^2_A \:(\mathrm{fm}^2)$\\ \hline 
				No deuteron  & 24.4 & 1.17 $\pm$ 0.03 & 1000 $\pm$ 70  & 0.47 $\pm$ 0.07\\ \hline
				Deuteron  & 22.3 & 1.13 $\pm$ 0.03 & 950 $\pm$ 70  & 0.52 $\pm$ 0.08 \\   
				\hline
				\multicolumn{5}{|c|}{}\\
				\multicolumn{5}{|c|}{BIN2}\\
				\hline
				& $\chi^2$ & $p$ & $M_A \:(\mathrm{MeV})$  & $r^2_A \:(\mathrm{fm}^2)$\\ \hline 
				No deuteron  & 20.8 & 1.22 $\pm$ 0.05 & 890 $\pm$ 100  & 0.59 $\pm$ 0.13\\ \hline
				Deuteron  & 19.8 & 1.18 $\pm$ 0.05 & 850 $\pm$ 110  & 0.65 $\pm$ 0.16\\    \hline
			\end{tabular}}
			\caption{Fits of the dipole axial form factor (\ref{Eq:FA_dipole}) to the BIN0, BIN1 and BIN2 data sets.
				\label{Table_Minuit_ANL}}
		\end{center}
	\end{table}
	
Let us stress that the $r_A^2$ and normalization parameter $p$ for the fits of BIN0 data are comparable to the {\it z} expansion results~\cite{Meyer:2016oeg}, even though in the latter case, a different error function was utilized.  Certainly, with a dipole fit to BIN0 data one can not obtain the local maximum of $F_A$ at low $Q^2$. On the other hand, the dipole fits to the BIN1 and the BIN2 data have very similar functional $Q^2$ dependence as the best MLP fits. 
For these data sets, the evidence, which contains Occam's factor penalizing overfitting parametrizations with large error bars, establishes the preference for a rather structureless neural network that departs very little from the normalization values, given by Eq.~\ref{Eq:FA_dipole} with $M_A = 1$~GeV, and has small errors. Furthermore, the uncertainties in the neural-network fits are systematically smaller than in the dipole $\chi^2$ ones.    

\section{Summary}
\label{Sec:Summary}

The first  Bayesian neural-network analysis 
of the neutrino-deuteron scattering data has been performed. 
The reported study has been oriented to the extraction of the axial form factor from the ANL CCQE data, searching for deviations from the dipole form. With the full ANL data set, the analysis leads to an axial form factor which has a positive slope at $Q^2=0$ and a local maximum at low $Q^2$. The inclusion of the deuteron correction reduces the peak in $F_A$. Only after removing the lowest available $Q^2$ region ($0.05 < Q^2 < 0.10$~GeV$^2$) from the data, a value of the axial radius consistent with available determinations could be obtained. This suggests that corrections from the deuteron structure play a crucial role at low $Q^2$ but it could also be the case that the experimental errors in this kinematic region were underestimated. Analyses without the low-$Q^2$ data do not show any significant deviation from 
previous determinations. Furthermore, our neural-network fits are characterized by smaller uncertainties than the dipole ones.

New more precise measurements of the neutrino cross sections on hydrogen and deuterium are needed to unravel the axial structure of the nucleon. Techniques, such as the one applied in the present paper will prove valuable in such a scenario.

\begin{acknowledgments}
We thank A. Meyer and M. Gonz\'alez-Alonso for useful communications. The calculations have been carried out in the Wroclaw Centre for Networking and Supercomputing (Ref.~\cite{Wroclaw}),
Grant No. 268. This work has been partially supported by the Spanish Ministerio de Econ\-om\'ia y Competitividad and the European Regional Development Fund under Contracts No. FIS2014-51948-C2-1-P, No. FIS2014-51948-C2-2-P, No. FIS2017-84038-C2-1-P, No. FIS2017-84038-C2-1-P and No. SEV-2014-0398 and by Generalitat Valenciana under Contract No. PROMETEOII/2014/0068. 
\end{acknowledgments}


\appendix

\section{Prior distributions}
\label{Appeendix_Prior}

The prior distribution of the weights, Eqs.~(\ref{Eq:Prior_Ew} and \ref{Eq:Eq}), is justified 
by the following properties of the adopted feed-forward neural network in the MLP configuration and of the problem under study
\begin{enumerate}[(P1)]

\item \label{Property_3b} internal symmetry: The exchange of any two units in the hidden layer does not change the functional form of the network and its output values;

\item \label{Property_2} the sigmoid activation function $f(x)$ Eq.~(\ref{Eq:sigmoid}) saturates and can be effectively assumed to be constant outside the interval $-a \leq x \leq a$ with $a \sim 10$;

\item \label{Property_3} $f(-x) = 1 - f(x)$; 

\item \label{Property_1} the ANL data are concentrated in the region $Q^2\in (0,3)$~GeV$^2$ -- constraint (\ref{assumption1}) in  Sec. \ref{Sec_Axial_form_factor};

 \item \label{Property_4} $F_A(Q^2)/F_A^{\mathrm{dipole}}(Q^2) < C$, where $C>1$ --  constraint (\ref{assumption4}) in Sec. \ref{Sec_Axial_form_factor}.
 \end{enumerate}

Properties (P\ref{Property_1}) and (P\ref{Property_2}) constrain  the weights $w_1,...,w_{2M}$ in function (\ref{Eq:functional:form:of:MLP}). Indeed, from  (P\ref{Property_2}) one sees  that for a full performance of the activation function $f(x)$ in Eq.~(\ref{Eq:functional:form:of:MLP}) it is enough to have $x \in (-a,a)$. 
Let us consider then the function $f(w_i Q^2 +w_{M+i}) $, $i=1 \ldots, M$, which  is  one of the elements of the sum  (\ref{Eq:functional:form:of:MLP}). 
To efficiently cover all possible outputs, achieving a good performance of the network, 
it is enough that the arguments of $f(...)$ belong to $(-a,a)$ for all values of $Q^2$ under consideration. Then, one gets limits for the weights, namely, $|w_{M+i} | < a$ \footnote{At $Q^2=0$ we have $f(w_{M+i})$, so it is enough that $|w_{M+i}|<a$.} from which one also gets the constraint  $|w_{i}| < 2/3 a < a$ for $Q^2  \in (0,3)$~GeV$^2$.

Note that both negative and positive values of weights $w_{i}$ and $w_{M+i}$ are equally 
possible according to (P\ref{Property_3}). Therefore, without losing generality, the prior density should be symmetric in weights $w_{1 \mbox{-} 2M}$ and cover the hypercube $(-a,a)^{2M}$.

The limits for the weights in the linear layer are less obvious. Property (P\ref{Property_4}) provides a constraint on the weights in the linear output layer $w_{2M+i}$, $i=1, \dots, M+1$, namely, 
\begin{eqnarray}
\left|\sum_{i=1}^M w_{2M+i} f(\dots) + {w}_{3M+1}\right|  < C\,.
\end{eqnarray}
The main role of the weights in the linear layer is to control the range of the neural-network output. In our case, at any $Q^2$, the absolute value of the output should be smaller than $C$. Hence, for a reliable performance of the network, it is enough to assume that the weights in the linear layer are $|w_{2M+i}| < C \sim a$, $i=1, \cdots, M+1$. In analogy to the reasoning in the paragraph above, one can argue that these weights could equally take positive and negative values.

We, therefore, conclude that the prior density for the weights should cover the hypercube $(-a,a)^{3M+1}$ and be symmetric in $w_i$.
This is a rough estimate of the bounds for the model parameters, but the functional form of the prior densities is still arbitrary. In the present analysis, we have considered the Gaussian distribution (\ref{Eq:Prior_Ew}). Such a density profile maximizes the entropy of the system \cite{Lepage:2001ym}.  However, our choice is also supported by further arguments from  neural-network computations. In particular, prior (\ref{Eq:Prior_Ew}) modifies the error function (\ref{Eq:Error_function}) through the  contribution (\ref{Eq:Eq}). Such a penalty term has been considered in non-Bayesian approaches~\cite{10.1007/3-540-17943-7_117} to the feed forward neural networks. Bayesian statistics provides a probabilistic justification for (\ref{Eq:Eq}). Moreover, the  Bayesian approach allows to consistently obtain the optimal value of the $\alpha$. We recall that in the numerical analysis, the initial value of $\alpha$ is set to $\alpha_0=0.001$. Then, the prior Gaussian distribution has a width of $\sqrt{1/\alpha} \approx 30 $, which fully covers the region of the parameter space
allowed by the constraints.

Eventually, let us remark that the most general Gaussian prior has the form 
\begin{equation}
P(\{w_i\}\mid \mathcal{N}) = 
\exp\left(- \sum_{i=1}^{3M+1} \frac{\alpha_i}{2} w_i^2\right)\,,
\end{equation}
where every weight $w_i$ has its own regularizer $\alpha_i$. However, the internal symmetry of the network (P\ref{Property_3b}) allows to reduce the number of regularizers to only four -- each for every class of parameters ($w_{i}$, $w_{M+i}$, $w_{2M+i}$, $w_{3M+1}$) with $i=1-M$. In Ref. \cite{Graczyk:2010gw}, it was verified that the inclusion of more regularizers has a negligible impact on the results but slows down the numerical procedures. Hence, in the present analysis, we consider the simplest and practically very effective scenario.

\begin{widetext}
\section{Best-fit results}
\label{App:Fits}

It is worth noting that each of the sigmoids that constitute the neural networks typically describes a  particular feature of the function. If a soft dependence is preferred by the data, some units might be redundant
and take very similar values for the weights.

\subsection{BIN0 with deuteron corrections}
 
 The best-fit parametrization for the BIN0 data set with the deuteron correction included is 
 \begin{eqnarray} 
 \mathcal{N}(Q^2,\{w_j\})
 &=&\frac{w_9}{e^{ -\frac{Q^2}{Q_0^2} w_1-w_2}+1}+\frac{w_{10}}{e^{-\frac{Q^2}{Q_0^2} w_3-w_4}+1}+\frac{w_{11}}{e^{-\frac{Q^2}{Q_0^2} w_5-w_6}+1}+\frac{w_{12}}{e^{-\frac{Q^2}{Q_0^2} w_7-w_8}+1}+w_{13} \,.
 \end{eqnarray} 

The weights $w_{1-13}$ take the following values:
\begin{eqnarray} \{w_j\}&=&\{-2.174061,0.1991515,2.140942,-0.1947798,-2.174070,
 0.1991740,\nonumber \\
 & &
 -5.481409,2.501837,-2.502352,2.308397,-2.502347,3.120895,-0.1638095\}
\end{eqnarray}
with a covariance matrix 
 \begin{scriptsize}
 \begin{equation}
 \resizebox{0.99\hsize}{!}{$
 A^{-1} = \left(
\begin{array}{ccccccccccccc}
 3.037317 & -0.1350095 & 0.7164281 & 0.6028375 & -0.9106402 & -0.7254212 & -0.1375870 & 0.1193517 & 1.744848 & 0.7934970 & -1.638952 & -0.6857111 & -0.6863036 \\
 -0.1350095 & 6.839114 & 0.5963543 & 2.241853 & -0.7253642 & -2.521010 & 0.08826055 & 0.08952850 & 0.5380971 & -0.4953074 & 0.7737700 & 0.5333276 & 0.3832513 \\
 0.7164281 & 0.5963543 & 3.129677 & -0.3601146 & 0.7164453 & 0.5964068 & 0.08461544 & -0.05068235 & 1.066743 & 1.171111 & 1.066819 & 0.6044601 & -1.304295 \\
 0.6028375 & 2.241853 & -0.3601146 & 7.284083 & 0.6028388 & 2.242126 & -0.1002227 & -0.05207626 & -0.8354786 & 0.1749902 & -0.8355188 & -0.4839537 & -0.08842895 \\
 -0.9106402 & -0.7253642 & 0.7164453 & 0.6028388 & 3.037309 & -0.1350448 & -0.1375865 & 0.1193541 & -1.638966 & 0.7934769 & 1.744873 & -0.6857391 & -0.6862827 \\
 -0.7254212 & -2.521010 & 0.5964068 & 2.242126 & -0.1350448 & 6.839396 & 0.08828352 & 0.08953791 & 0.7736800 & -0.4953435 & 0.5381072 & 0.5333633 & 0.3832724 \\
 -0.1375870 & 0.08826055 & 0.08461544 & -0.1002227 & -0.1375865 & 0.08828352 & 1.348550 & -0.7524626 & -0.8319859 & 0.4259322 & -0.8320014 & 1.535086 & -0.3655335 \\
 0.1193517 & 0.08952850 & -0.05068235 & -0.05207627 & 0.1193541 & 0.08953791 & -0.7524626 & 0.5275705 & 0.6937163 & -0.3336553 & 0.6937285 & -0.9665196 & 0.3068238 \\
 1.744848 & 0.5380971 & 1.066743 & -0.8354786 & -1.638966 & 0.7736800 & -0.8319859 & 0.6937163 & 13.95026 & 2.966996 & -6.650710 & -1.081263 & -3.213450 \\
 0.7934970 & -0.4953074 & 1.171111 & 0.1749902 & 0.7934769 & -0.4953435 & 0.4259322 & -0.3336553 & 2.966996 & 8.333349 & 2.967110 & 0.5435698 & -8.166890 \\
 -1.638952 & 0.7737700 & 1.066819 & -0.8355188 & 1.744873 & 0.5381072 & -0.8320014 & 0.6937285 & -6.650710 & 2.967110 & 13.95042 & -1.081328 & -3.213562 \\
 -0.6857111 & 0.5333276 & 0.6044601 & -0.4839537 & -0.6857391 & 0.5333633 & 1.535086 & -0.9665196 & -1.081263 & 0.5435698 & -1.081328 & 2.761672 & -0.4607885 \\
 -0.6863036 & 0.3832513 & -1.304295 & -0.08842895 & -0.6862827 & 0.3832724 & -0.3655335 & 0.3068238 & -3.213450 & -8.166890 & -3.213562 & -0.4607885 & 8.103694 \\
\end{array}
\right) $}\end{equation}
\end{scriptsize}
As explained in Sec.~\ref{Sec:Bayesian:sub:MLP}, Eq.~\ref{Eq:FA_NN}, to obtain $F_A(Q^2)$, function $\mathcal{N}(Q^2,\{w_j\})$ given above should be multiplied by the dipole Eq.~\ref{Eq:FA_dipole} with $M_A = 1$~GeV.

\subsection{BIN1 with deuteron corrections}

In this case,
\begin{eqnarray} 
 \mathcal{N}(Q^2,\{w_j\})&=& 
 \frac{w_7}{e^{-\frac{Q^2}{Q_0^2} w_1- w_2}+1}+\frac{w_8}{e^{-\frac{Q^2}{Q_0^2} w_3-w_4}+1}+\frac{w_9}{e^{-\frac{Q^2}{Q_0^2} w_5-w_6}+1}+w_{10} \,, 
 \end{eqnarray} 
with 
\begin{equation} \{w_j\}_{MP}=\{-0.0703401,0.0404197,-0.0703404,0.0404186,-0.0703372,0.0404192,0.299085,0.299087,0.299086,0.554479\} \,,
\end{equation}
and a  covariance matrix 
\begin{scriptsize}
 \begin{equation} A^{-1} = \left(
\begin{array}{cccccccccc}
 0.501347 & 0.000475574 & -0.133756 & 0.00312538 & -0.133756 & 0.00312517 & -0.0951767 & 0.0545460 & 0.0545442 & -0.00759898 \\
 0.000475574 & 0.563049 & 0.00312506 & -0.00641521 & 0.00312465 & -0.00641482 & 0.0538884 & -0.0236924 & -0.0236923 & -0.0444412 \\
 -0.133756 & 0.00312506 & 0.501346 & 0.000475531 & -0.133757 & 0.00312475 & 0.0545453 & -0.0951765 & 0.0545448 & -0.00759898 \\
 0.00312538 & -0.00641521 & 0.000475531 & 0.563050 & 0.00312457 & -0.00641496 & -0.0236928 & 0.0538882 & -0.0236918 & -0.0444415 \\
 -0.133756 & 0.00312465 & -0.133757 & 0.00312457 & 0.501347 & 0.000474719 & 0.0545458 & 0.0545457 & -0.0951783 & -0.00759868 \\
 0.00312517 & -0.00641482 & 0.00312475 & -0.00641496 & 0.000474719 & 0.563050 & -0.0236928 & -0.0236921 & 0.0538888 & -0.0444415 \\
 -0.0951767 & 0.0538884 & 0.0545453 & -0.0236928 & 0.0545458 & -0.0236928 & 0.505143 & -0.101488 & -0.101488 & -0.154622 \\
 0.0545460 & -0.0236924 & -0.0951765 & 0.0538882 & 0.0545457 & -0.0236921 & -0.101488 & 0.505144 & -0.101488 & -0.154623 \\
 0.0545442 & -0.0236923 & 0.0545448 & -0.0236918 & -0.0951783 & 0.0538888 & -0.101488 & -0.101488 & 0.505144 & -0.154623 \\
 -0.00759898 & -0.0444412 & -0.00759898 & -0.0444415 & -0.00759868 & -0.0444415 & -0.154622 & -0.154623 & -0.154623 & 0.246587 \\
\end{array}
\right) \,. \end{equation}
\end{scriptsize}

\end{widetext}

\bibliography{bibliography}

\begin{thebibliography}{64}%
\makeatletter
\providecommand \@ifxundefined [1]{%
 \@ifx{#1\undefined}
}%
\providecommand \@ifnum [1]{%
 \ifnum #1\expandafter \@firstoftwo
 \else \expandafter \@secondoftwo
 \fi
}%
\providecommand \@ifx [1]{%
 \ifx #1\expandafter \@firstoftwo
 \else \expandafter \@secondoftwo
 \fi
}%
\providecommand \natexlab [1]{#1}%
\providecommand \enquote  [1]{``#1''}%
\providecommand \bibnamefont  [1]{#1}%
\providecommand \bibfnamefont [1]{#1}%
\providecommand \citenamefont [1]{#1}%
\providecommand \href@noop [0]{\@secondoftwo}%
\providecommand \href [0]{\begingroup \@sanitize@url \@href}%
\providecommand \@href[1]{\@@startlink{#1}\@@href}%
\providecommand \@@href[1]{\endgroup#1\@@endlink}%
\providecommand \@sanitize@url [0]{\catcode `\\12\catcode `\$12\catcode
  `\&12\catcode `\#12\catcode `\^12\catcode `\_12\catcode `\%12\relax}%
\providecommand \@@startlink[1]{}%
\providecommand \@@endlink[0]{}%
\providecommand \url  [0]{\begingroup\@sanitize@url \@url }%
\providecommand \@url [1]{\endgroup\@href {#1}{\urlprefix }}%
\providecommand \urlprefix  [0]{URL }%
\providecommand \Eprint [0]{\href }%
\providecommand \doibase [0]{http://dx.doi.org/}%
\providecommand \selectlanguage [0]{\@gobble}%
\providecommand \bibinfo  [0]{\@secondoftwo}%
\providecommand \bibfield  [0]{\@secondoftwo}%
\providecommand \translation [1]{[#1]}%
\providecommand \BibitemOpen [0]{}%
\providecommand \bibitemStop [0]{}%
\providecommand \bibitemNoStop [0]{.\EOS\space}%
\providecommand \EOS [0]{\spacefactor3000\relax}%
\providecommand \BibitemShut  [1]{\csname bibitem#1\endcsname}%
\let\auto@bib@innerbib\@empty
\bibitem [{\citenamefont {Alvarez-Ruso}\ \emph {et~al.}(2017)\citenamefont
  {Alvarez-Ruso} \emph {et~al.}}]{Alvarez-Ruso:2017oui}%
  \BibitemOpen
  \bibfield  {author} {\bibinfo {author} {\bibfnamefont {L.}~\bibnamefont
  {Alvarez-Ruso}} \emph {et~al.},\ }\href@noop {} {\  (\bibinfo {year}
  {2017})},\ \Eprint {http://arxiv.org/abs/1706.03621} {arXiv:1706.03621
  [hep-ph]} \BibitemShut {NoStop}%
\bibitem [{\citenamefont {Patrignani}\ \emph {et~al.}(2016)\citenamefont
  {Patrignani} \emph {et~al.}}]{Patrignani:2016xqp}%
  \BibitemOpen
  \bibfield  {author} {\bibinfo {author} {\bibfnamefont {C.}~\bibnamefont
  {Patrignani}} \emph {et~al.} (\bibinfo {collaboration} {Particle Data
  Group}),\ }\href {\doibase 10.1088/1674-1137/40/10/100001} {\bibfield
  {journal} {\bibinfo  {journal} {Chin. Phys.}\ }\textbf {\bibinfo {volume}
  {C40}},\ \bibinfo {pages} {100001} (\bibinfo {year} {2016})}\BibitemShut
  {NoStop}%
\bibitem [{\citenamefont {Gonz\'alez-Alonso}\ \emph {et~al.}(2018)\citenamefont
  {Gonz\'alez-Alonso}, \citenamefont {Naviliat-Cuncic},\ and\ \citenamefont
  {Severijns}}]{Gonzalez-Alonso:2018omy}%
  \BibitemOpen
  \bibfield  {author} {\bibinfo {author} {\bibfnamefont {M.}~\bibnamefont
  {Gonz\'alez-Alonso}}, \bibinfo {author} {\bibfnamefont {O.}~\bibnamefont
  {Naviliat-Cuncic}}, \ and\ \bibinfo {author} {\bibfnamefont {N.}~\bibnamefont
  {Severijns}},\ }\href@noop {} {\  (\bibinfo {year} {2018})},\ \Eprint
  {http://arxiv.org/abs/1803.08732} {arXiv:1803.08732 [hep-ph]} \BibitemShut
  {NoStop}%
\bibitem [{\citenamefont {Arrington}\ \emph {et~al.}(2007)\citenamefont
  {Arrington}, \citenamefont {Roberts},\ and\ \citenamefont
  {Zanotti}}]{Arrington:2006zm}%
  \BibitemOpen
  \bibfield  {author} {\bibinfo {author} {\bibfnamefont {J.}~\bibnamefont
  {Arrington}}, \bibinfo {author} {\bibfnamefont {C.~D.}\ \bibnamefont
  {Roberts}}, \ and\ \bibinfo {author} {\bibfnamefont {J.~M.}\ \bibnamefont
  {Zanotti}},\ }\href {\doibase 10.1088/0954-3899/34/7/S03} {\bibfield
  {journal} {\bibinfo  {journal} {J. Phys.}\ }\textbf {\bibinfo {volume}
  {G34}},\ \bibinfo {pages} {S23} (\bibinfo {year} {2007})},\ \Eprint
  {http://arxiv.org/abs/nucl-th/0611050} {arXiv:nucl-th/0611050 [nucl-th]}
  \BibitemShut {NoStop}%
\bibitem [{\citenamefont {Katori}\ and\ \citenamefont
  {Martini}(2018)}]{Katori:2016yel}%
  \BibitemOpen
  \bibfield  {author} {\bibinfo {author} {\bibfnamefont {T.}~\bibnamefont
  {Katori}}\ and\ \bibinfo {author} {\bibfnamefont {M.}~\bibnamefont
  {Martini}},\ }\href {\doibase 10.1088/1361-6471/aa8bf7} {\bibfield  {journal}
  {\bibinfo  {journal} {J. Phys.}\ }\textbf {\bibinfo {volume} {G45}},\
  \bibinfo {pages} {013001} (\bibinfo {year} {2018})},\ \Eprint
  {http://arxiv.org/abs/1611.07770} {arXiv:1611.07770 [hep-ph]} \BibitemShut
  {NoStop}%
\bibitem [{\citenamefont {Mann}\ \emph {et~al.}(1973)\citenamefont {Mann} \emph
  {et~al.}}]{Mann:1973pr}%
  \BibitemOpen
  \bibfield  {author} {\bibinfo {author} {\bibfnamefont {W.~A.}\ \bibnamefont
  {Mann}} \emph {et~al.},\ }\href {\doibase 10.1103/PhysRevLett.31.844}
  {\bibfield  {journal} {\bibinfo  {journal} {Phys. Rev. Lett.}\ }\textbf
  {\bibinfo {volume} {31}},\ \bibinfo {pages} {844} (\bibinfo {year}
  {1973})}\BibitemShut {NoStop}%
\bibitem [{\citenamefont {Barish}\ \emph {et~al.}(1977)\citenamefont {Barish}
  \emph {et~al.}}]{Barish:1977qk}%
  \BibitemOpen
  \bibfield  {author} {\bibinfo {author} {\bibfnamefont {S.~J.}\ \bibnamefont
  {Barish}} \emph {et~al.},\ }\href {\doibase 10.1103/PhysRevD.16.3103}
  {\bibfield  {journal} {\bibinfo  {journal} {Phys. Rev.}\ }\textbf {\bibinfo
  {volume} {D16}},\ \bibinfo {pages} {3103} (\bibinfo {year}
  {1977})}\BibitemShut {NoStop}%
\bibitem [{\citenamefont {Miller}\ \emph {et~al.}(1982)\citenamefont {Miller}
  \emph {et~al.}}]{Miller:1982qi}%
  \BibitemOpen
  \bibfield  {author} {\bibinfo {author} {\bibfnamefont {K.~L.}\ \bibnamefont
  {Miller}} \emph {et~al.},\ }\href {\doibase 10.1103/PhysRevD.26.537}
  {\bibfield  {journal} {\bibinfo  {journal} {Phys. Rev.}\ }\textbf {\bibinfo
  {volume} {D26}},\ \bibinfo {pages} {537} (\bibinfo {year}
  {1982})}\BibitemShut {NoStop}%
\bibitem [{\citenamefont {Baker}\ \emph {et~al.}(1981)\citenamefont {Baker},
  \citenamefont {Cnops}, \citenamefont {Connolly}, \citenamefont {Kahn},
  \citenamefont {Kirk}, \citenamefont {Murtagh}, \citenamefont {Palmer},
  \citenamefont {Samios},\ and\ \citenamefont {Tanaka}}]{Baker:1981su}%
  \BibitemOpen
  \bibfield  {author} {\bibinfo {author} {\bibfnamefont {N.~J.}\ \bibnamefont
  {Baker}}, \bibinfo {author} {\bibfnamefont {A.~M.}\ \bibnamefont {Cnops}},
  \bibinfo {author} {\bibfnamefont {P.~L.}\ \bibnamefont {Connolly}}, \bibinfo
  {author} {\bibfnamefont {S.~A.}\ \bibnamefont {Kahn}}, \bibinfo {author}
  {\bibfnamefont {H.~G.}\ \bibnamefont {Kirk}}, \bibinfo {author}
  {\bibfnamefont {M.~J.}\ \bibnamefont {Murtagh}}, \bibinfo {author}
  {\bibfnamefont {R.~B.}\ \bibnamefont {Palmer}}, \bibinfo {author}
  {\bibfnamefont {N.~P.}\ \bibnamefont {Samios}}, \ and\ \bibinfo {author}
  {\bibfnamefont {M.}~\bibnamefont {Tanaka}},\ }\href {\doibase
  10.1103/PhysRevD.23.2499} {\bibfield  {journal} {\bibinfo  {journal} {Phys.
  Rev.}\ }\textbf {\bibinfo {volume} {D23}},\ \bibinfo {pages} {2499} (\bibinfo
  {year} {1981})}\BibitemShut {NoStop}%
\bibitem [{\citenamefont {Kitagaki}\ \emph {et~al.}(1990)\citenamefont
  {Kitagaki} \emph {et~al.}}]{Kitagaki:1990vs}%
  \BibitemOpen
  \bibfield  {author} {\bibinfo {author} {\bibfnamefont {T.}~\bibnamefont
  {Kitagaki}} \emph {et~al.},\ }\href {\doibase 10.1103/PhysRevD.42.1331}
  {\bibfield  {journal} {\bibinfo  {journal} {Phys. Rev.}\ }\textbf {\bibinfo
  {volume} {D42}},\ \bibinfo {pages} {1331} (\bibinfo {year}
  {1990})}\BibitemShut {NoStop}%
\bibitem [{\citenamefont {Kitagaki}\ \emph {et~al.}(1983)\citenamefont
  {Kitagaki} \emph {et~al.}}]{Kitagaki:1983px}%
  \BibitemOpen
  \bibfield  {author} {\bibinfo {author} {\bibfnamefont {T.}~\bibnamefont
  {Kitagaki}} \emph {et~al.},\ }\href {\doibase 10.1103/PhysRevD.28.436}
  {\bibfield  {journal} {\bibinfo  {journal} {Phys. Rev.}\ }\textbf {\bibinfo
  {volume} {D28}},\ \bibinfo {pages} {436} (\bibinfo {year}
  {1983})}\BibitemShut {NoStop}%
\bibitem [{\citenamefont {Allasia}\ \emph {et~al.}(1990)\citenamefont {Allasia}
  \emph {et~al.}}]{Allasia:1990uy}%
  \BibitemOpen
  \bibfield  {author} {\bibinfo {author} {\bibfnamefont {D.}~\bibnamefont
  {Allasia}} \emph {et~al.},\ }\href {\doibase 10.1016/0550-3213(90)90472-P}
  {\bibfield  {journal} {\bibinfo  {journal} {Nucl. Phys.}\ }\textbf {\bibinfo
  {volume} {B343}},\ \bibinfo {pages} {285} (\bibinfo {year}
  {1990})}\BibitemShut {NoStop}%
\bibitem [{\citenamefont {Bodek}\ \emph {et~al.}(2008)\citenamefont {Bodek},
  \citenamefont {Avvakumov}, \citenamefont {Bradford},\ and\ \citenamefont
  {Budd}}]{Bodek:2007ym}%
  \BibitemOpen
  \bibfield  {author} {\bibinfo {author} {\bibfnamefont {A.}~\bibnamefont
  {Bodek}}, \bibinfo {author} {\bibfnamefont {S.}~\bibnamefont {Avvakumov}},
  \bibinfo {author} {\bibfnamefont {R.}~\bibnamefont {Bradford}}, \ and\
  \bibinfo {author} {\bibfnamefont {H.~S.}\ \bibnamefont {Budd}},\ }\href
  {\doibase 10.1140/epjc/s10052-007-0491-4} {\bibfield  {journal} {\bibinfo
  {journal} {Eur. Phys. J.}\ }\textbf {\bibinfo {volume} {C53}},\ \bibinfo
  {pages} {349} (\bibinfo {year} {2008})},\ \Eprint
  {http://arxiv.org/abs/0708.1946} {arXiv:0708.1946 [hep-ex]} \BibitemShut
  {NoStop}%
\bibitem [{\citenamefont {Bhattacharya}\ \emph {et~al.}(2011)\citenamefont
  {Bhattacharya}, \citenamefont {Hill},\ and\ \citenamefont
  {Paz}}]{Bhattacharya:2011ah}%
  \BibitemOpen
  \bibfield  {author} {\bibinfo {author} {\bibfnamefont {B.}~\bibnamefont
  {Bhattacharya}}, \bibinfo {author} {\bibfnamefont {R.~J.}\ \bibnamefont
  {Hill}}, \ and\ \bibinfo {author} {\bibfnamefont {G.}~\bibnamefont {Paz}},\
  }\href {\doibase 10.1103/PhysRevD.84.073006} {\bibfield  {journal} {\bibinfo
  {journal} {Phys. Rev.}\ }\textbf {\bibinfo {volume} {D84}},\ \bibinfo {pages}
  {073006} (\bibinfo {year} {2011})},\ \Eprint {http://arxiv.org/abs/1108.0423}
  {arXiv:1108.0423 [hep-ph]} \BibitemShut {NoStop}%
\bibitem [{\citenamefont {Bhattacharya}\ \emph {et~al.}(2015)\citenamefont
  {Bhattacharya}, \citenamefont {Paz},\ and\ \citenamefont
  {Tropiano}}]{Bhattacharya:2015mpa}%
  \BibitemOpen
  \bibfield  {author} {\bibinfo {author} {\bibfnamefont {B.}~\bibnamefont
  {Bhattacharya}}, \bibinfo {author} {\bibfnamefont {G.}~\bibnamefont {Paz}}, \
  and\ \bibinfo {author} {\bibfnamefont {A.~J.}\ \bibnamefont {Tropiano}},\
  }\href {\doibase 10.1103/PhysRevD.92.113011} {\bibfield  {journal} {\bibinfo
  {journal} {Phys. Rev.}\ }\textbf {\bibinfo {volume} {D92}},\ \bibinfo {pages}
  {113011} (\bibinfo {year} {2015})},\ \Eprint
  {http://arxiv.org/abs/1510.05652} {arXiv:1510.05652 [hep-ph]} \BibitemShut
  {NoStop}%
\bibitem [{\citenamefont {Meyer}\ \emph {et~al.}(2016)\citenamefont {Meyer},
  \citenamefont {Betancourt}, \citenamefont {Gran},\ and\ \citenamefont
  {Hill}}]{Meyer:2016oeg}%
  \BibitemOpen
  \bibfield  {author} {\bibinfo {author} {\bibfnamefont {A.~S.}\ \bibnamefont
  {Meyer}}, \bibinfo {author} {\bibfnamefont {M.}~\bibnamefont {Betancourt}},
  \bibinfo {author} {\bibfnamefont {R.}~\bibnamefont {Gran}}, \ and\ \bibinfo
  {author} {\bibfnamefont {R.~J.}\ \bibnamefont {Hill}},\ }\href {\doibase
  10.1103/PhysRevD.93.113015} {\bibfield  {journal} {\bibinfo  {journal} {Phys.
  Rev.}\ }\textbf {\bibinfo {volume} {D93}},\ \bibinfo {pages} {113015}
  (\bibinfo {year} {2016})},\ \Eprint {http://arxiv.org/abs/1603.03048}
  {arXiv:1603.03048 [hep-ph]} \BibitemShut {NoStop}%
\bibitem [{\citenamefont {Hill}\ \emph {et~al.}(2018)\citenamefont {Hill},
  \citenamefont {Kammel}, \citenamefont {Marciano},\ and\ \citenamefont
  {Sirlin}}]{Hill:2017wgb}%
  \BibitemOpen
  \bibfield  {author} {\bibinfo {author} {\bibfnamefont {R.~J.}\ \bibnamefont
  {Hill}}, \bibinfo {author} {\bibfnamefont {P.}~\bibnamefont {Kammel}},
  \bibinfo {author} {\bibfnamefont {W.~J.}\ \bibnamefont {Marciano}}, \ and\
  \bibinfo {author} {\bibfnamefont {A.}~\bibnamefont {Sirlin}},\ }\href
  {\doibase 10.1088/1361-6633/aac190} {\bibfield  {journal} {\bibinfo
  {journal} {Rept. Prog. Phys.}\ }\textbf {\bibinfo {volume} {81}},\ \bibinfo
  {pages} {096301} (\bibinfo {year} {2018})},\ \Eprint
  {http://arxiv.org/abs/1708.08462} {arXiv:1708.08462 [hep-ph]} \BibitemShut
  {NoStop}%
\bibitem [{\citenamefont {Berkowitz}\ \emph {et~al.}(2017)\citenamefont
  {Berkowitz} \emph {et~al.}}]{Berkowitz:2017gql}%
  \BibitemOpen
  \bibfield  {author} {\bibinfo {author} {\bibfnamefont {E.}~\bibnamefont
  {Berkowitz}} \emph {et~al.},\ }\href@noop {} {\  (\bibinfo {year} {2017})},\
  \Eprint {http://arxiv.org/abs/1704.01114} {arXiv:1704.01114 [hep-lat]}
  \BibitemShut {NoStop}%
\bibitem [{\citenamefont {Alexandrou}\ \emph {et~al.}(2017)\citenamefont
  {Alexandrou}, \citenamefont {Constantinou}, \citenamefont {Hadjiyiannakou},
  \citenamefont {Jansen}, \citenamefont {Kallidonis}, \citenamefont {Koutsou},\
  and\ \citenamefont {Vaquero Aviles-Casco}}]{Alexandrou:2017hac}%
  \BibitemOpen
  \bibfield  {author} {\bibinfo {author} {\bibfnamefont {C.}~\bibnamefont
  {Alexandrou}}, \bibinfo {author} {\bibfnamefont {M.}~\bibnamefont
  {Constantinou}}, \bibinfo {author} {\bibfnamefont {K.}~\bibnamefont
  {Hadjiyiannakou}}, \bibinfo {author} {\bibfnamefont {K.}~\bibnamefont
  {Jansen}}, \bibinfo {author} {\bibfnamefont {C.}~\bibnamefont {Kallidonis}},
  \bibinfo {author} {\bibfnamefont {G.}~\bibnamefont {Koutsou}}, \ and\
  \bibinfo {author} {\bibfnamefont {A.}~\bibnamefont {Vaquero Aviles-Casco}},\
  }\href {\doibase 10.1103/PhysRevD.96.054507} {\bibfield  {journal} {\bibinfo
  {journal} {Phys. Rev.}\ }\textbf {\bibinfo {volume} {D96}},\ \bibinfo {pages}
  {054507} (\bibinfo {year} {2017})},\ \Eprint
  {http://arxiv.org/abs/1705.03399} {arXiv:1705.03399 [hep-lat]} \BibitemShut
  {NoStop}%
\bibitem [{\citenamefont {Capitani}\ \emph {et~al.}(2017)\citenamefont
  {Capitani}, \citenamefont {Della~Morte}, \citenamefont {Djukanovic},
  \citenamefont {von Hippel}, \citenamefont {Hua}, \citenamefont {Jäger},
  \citenamefont {Junnarkar}, \citenamefont {Meyer}, \citenamefont {Rae},\ and\
  \citenamefont {Wittig}}]{Capitani:2017qpc}%
  \BibitemOpen
  \bibfield  {author} {\bibinfo {author} {\bibfnamefont {S.}~\bibnamefont
  {Capitani}}, \bibinfo {author} {\bibfnamefont {M.}~\bibnamefont
  {Della~Morte}}, \bibinfo {author} {\bibfnamefont {D.}~\bibnamefont
  {Djukanovic}}, \bibinfo {author} {\bibfnamefont {G.~M.}\ \bibnamefont {von
  Hippel}}, \bibinfo {author} {\bibfnamefont {J.}~\bibnamefont {Hua}}, \bibinfo
  {author} {\bibfnamefont {B.}~\bibnamefont {Jäger}}, \bibinfo {author}
  {\bibfnamefont {P.~M.}\ \bibnamefont {Junnarkar}}, \bibinfo {author}
  {\bibfnamefont {H.~B.}\ \bibnamefont {Meyer}}, \bibinfo {author}
  {\bibfnamefont {T.~D.}\ \bibnamefont {Rae}}, \ and\ \bibinfo {author}
  {\bibfnamefont {H.}~\bibnamefont {Wittig}},\ }\href@noop {} {\  (\bibinfo
  {year} {2017})},\ \Eprint {http://arxiv.org/abs/1705.06186} {arXiv:1705.06186
  [hep-lat]} \BibitemShut {NoStop}%
\bibitem [{\citenamefont {Gupta}\ \emph {et~al.}(2017)\citenamefont {Gupta},
  \citenamefont {Jang}, \citenamefont {Lin}, \citenamefont {Yoon},\ and\
  \citenamefont {Bhattacharya}}]{Rajan:2017lxk}%
  \BibitemOpen
  \bibfield  {author} {\bibinfo {author} {\bibfnamefont {R.}~\bibnamefont
  {Gupta}}, \bibinfo {author} {\bibfnamefont {Y.-C.}\ \bibnamefont {Jang}},
  \bibinfo {author} {\bibfnamefont {H.-W.}\ \bibnamefont {Lin}}, \bibinfo
  {author} {\bibfnamefont {B.}~\bibnamefont {Yoon}}, \ and\ \bibinfo {author}
  {\bibfnamefont {T.}~\bibnamefont {Bhattacharya}},\ }\href {\doibase
  10.1103/PhysRevD.96.114503} {\bibfield  {journal} {\bibinfo  {journal} {Phys.
  Rev.}\ }\textbf {\bibinfo {volume} {D96}},\ \bibinfo {pages} {114503}
  (\bibinfo {year} {2017})},\ \Eprint {http://arxiv.org/abs/1705.06834}
  {arXiv:1705.06834 [hep-lat]} \BibitemShut {NoStop}%
\bibitem [{\citenamefont {Yao}\ \emph {et~al.}(2017)\citenamefont {Yao},
  \citenamefont {Alvarez-Ruso},\ and\ \citenamefont
  {Vicente-Vacas}}]{Yao:2017fym}%
  \BibitemOpen
  \bibfield  {author} {\bibinfo {author} {\bibfnamefont {D.-L.}\ \bibnamefont
  {Yao}}, \bibinfo {author} {\bibfnamefont {L.}~\bibnamefont {Alvarez-Ruso}}, \
  and\ \bibinfo {author} {\bibfnamefont {M.~J.}\ \bibnamefont
  {Vicente-Vacas}},\ }\href {\doibase 10.1103/PhysRevD.96.116022} {\bibfield
  {journal} {\bibinfo  {journal} {Phys. Rev.}\ }\textbf {\bibinfo {volume}
  {D96}},\ \bibinfo {pages} {116022} (\bibinfo {year} {2017})},\ \Eprint
  {http://arxiv.org/abs/1708.08776} {arXiv:1708.08776 [hep-ph]} \BibitemShut
  {NoStop}%
\bibitem [{\citenamefont {Graczyk}\ and\ \citenamefont
  {Juszczak}(2014)}]{Graczyk:2014lba}%
  \BibitemOpen
  \bibfield  {author} {\bibinfo {author} {\bibfnamefont {K.~M.}\ \bibnamefont
  {Graczyk}}\ and\ \bibinfo {author} {\bibfnamefont {C.}~\bibnamefont
  {Juszczak}},\ }\href {\doibase 10.1103/PhysRevC.90.054334} {\bibfield
  {journal} {\bibinfo  {journal} {Phys. Rev.}\ }\textbf {\bibinfo {volume}
  {C90}},\ \bibinfo {pages} {054334} (\bibinfo {year} {2014})},\ \Eprint
  {http://arxiv.org/abs/1408.0150} {arXiv:1408.0150 [hep-ph]} \BibitemShut
  {NoStop}%
\bibitem [{\citenamefont {Ball}\ \emph {et~al.}(2013)\citenamefont {Ball},
  \citenamefont {Forte}, \citenamefont {Guffanti}, \citenamefont {Nocera},
  \citenamefont {Ridolfi},\ and\ \citenamefont {Rojo}}]{Ball:2013lla}%
  \BibitemOpen
  \bibfield  {author} {\bibinfo {author} {\bibfnamefont {R.~D.}\ \bibnamefont
  {Ball}}, \bibinfo {author} {\bibfnamefont {S.}~\bibnamefont {Forte}},
  \bibinfo {author} {\bibfnamefont {A.}~\bibnamefont {Guffanti}}, \bibinfo
  {author} {\bibfnamefont {E.~R.}\ \bibnamefont {Nocera}}, \bibinfo {author}
  {\bibfnamefont {G.}~\bibnamefont {Ridolfi}}, \ and\ \bibinfo {author}
  {\bibfnamefont {J.}~\bibnamefont {Rojo}} (\bibinfo {collaboration} {NNPDF}),\
  }\href {\doibase 10.1016/j.nuclphysb.2013.05.007} {\bibfield  {journal}
  {\bibinfo  {journal} {Nucl. Phys.}\ }\textbf {\bibinfo {volume} {B874}},\
  \bibinfo {pages} {36} (\bibinfo {year} {2013})},\ \Eprint
  {http://arxiv.org/abs/1303.7236} {arXiv:1303.7236 [hep-ph]} \BibitemShut
  {NoStop}%
\bibitem [{\citenamefont {{C. M., Bishop}}(1995)}]{Bishop_book}%
  \BibitemOpen
  \bibfield  {author} {\bibinfo {author} {\bibnamefont {{C. M., Bishop}}},\
  }\href
  {https://global.oup.com/academic/product/neural-networks-for-pattern-recognition-9780198538646?q=Christopher%20Bishop&lang=en&cc=pl}
  {\emph {\bibinfo {title} {{Neural Networks for Pattern Recognition}}}}\
  (\bibinfo  {publisher} {Oxford University Press},\ \bibinfo {year}
  {1995})\BibitemShut {NoStop}%
\bibitem [{\citenamefont {Hornik}\ \emph {et~al.}(1989)\citenamefont {Hornik},
  \citenamefont {Sinchcombe},\ and\ \citenamefont {Halbert}}]{Hornik89}%
  \BibitemOpen
  \bibfield  {author} {\bibinfo {author} {\bibfnamefont {K.}~\bibnamefont
  {Hornik}}, \bibinfo {author} {\bibfnamefont {M.}~\bibnamefont {Sinchcombe}},
  \ and\ \bibinfo {author} {\bibfnamefont {W.}~\bibnamefont {Halbert}},\ }\href
  {\doibase http://www.sciencedirect.com/science/article/pii/0893608089900208}
  {\bibfield  {journal} {\bibinfo  {journal} {Neural Networks}\ }\textbf
  {\bibinfo {volume} {2}},\ \bibinfo {pages} {359} (\bibinfo {year}
  {1989})}\BibitemShut {NoStop}%
\bibitem [{\citenamefont {{G. D'Agostini}}(2003)}]{DAgostini_book}%
  \BibitemOpen
  \bibfield  {author} {\bibinfo {author} {\bibnamefont {{G. D'Agostini}}},\
  }\href {http://www.worldscientific.com/worldscibooks/10.1142/5262} {\emph
  {\bibinfo {title} {{Bayesian Reasoning in Data Analysis}}}}\ (\bibinfo
  {publisher} {World Scientific},\ \bibinfo {year} {2003})\BibitemShut
  {NoStop}%
\bibitem [{\citenamefont {De~Cruz}\ \emph {et~al.}(2012)\citenamefont
  {De~Cruz}, \citenamefont {Vrancx}, \citenamefont {Vancraeyveld},\ and\
  \citenamefont {Ryckebusch}}]{DeCruz:2011xi}%
  \BibitemOpen
  \bibfield  {author} {\bibinfo {author} {\bibfnamefont {L.}~\bibnamefont
  {De~Cruz}}, \bibinfo {author} {\bibfnamefont {T.}~\bibnamefont {Vrancx}},
  \bibinfo {author} {\bibfnamefont {P.}~\bibnamefont {Vancraeyveld}}, \ and\
  \bibinfo {author} {\bibfnamefont {J.}~\bibnamefont {Ryckebusch}},\ }\href
  {\doibase 10.1103/PhysRevLett.108.182002} {\bibfield  {journal} {\bibinfo
  {journal} {Phys. Rev. Lett.}\ }\textbf {\bibinfo {volume} {108}},\ \bibinfo
  {pages} {182002} (\bibinfo {year} {2012})},\ \Eprint
  {http://arxiv.org/abs/1111.6511} {arXiv:1111.6511 [nucl-th]} \BibitemShut
  {NoStop}%
\bibitem [{\citenamefont {McDonnell}\ \emph {et~al.}(2015)\citenamefont
  {McDonnell}, \citenamefont {Schunck}, \citenamefont {Higdon}, \citenamefont
  {Sarich}, \citenamefont {Wild},\ and\ \citenamefont
  {Nazarewicz}}]{McDonnell:2015sja}%
  \BibitemOpen
  \bibfield  {author} {\bibinfo {author} {\bibfnamefont {J.~D.}\ \bibnamefont
  {McDonnell}}, \bibinfo {author} {\bibfnamefont {N.}~\bibnamefont {Schunck}},
  \bibinfo {author} {\bibfnamefont {D.}~\bibnamefont {Higdon}}, \bibinfo
  {author} {\bibfnamefont {J.}~\bibnamefont {Sarich}}, \bibinfo {author}
  {\bibfnamefont {S.~M.}\ \bibnamefont {Wild}}, \ and\ \bibinfo {author}
  {\bibfnamefont {W.}~\bibnamefont {Nazarewicz}},\ }\href {\doibase
  10.1103/PhysRevLett.114.122501} {\bibfield  {journal} {\bibinfo  {journal}
  {Phys. Rev. Lett.}\ }\textbf {\bibinfo {volume} {114}},\ \bibinfo {pages}
  {122501} (\bibinfo {year} {2015})},\ \Eprint
  {http://arxiv.org/abs/1501.03572} {arXiv:1501.03572 [nucl-th]} \BibitemShut
  {NoStop}%
\bibitem [{\citenamefont {MacKay}(1991)}]{MacKay_thesis}%
  \BibitemOpen
  \bibfield  {author} {\bibinfo {author} {\bibfnamefont {D.}~\bibnamefont
  {MacKay}},\ }\emph {\bibinfo {title} {Bayesian Methods for Adaptive
  Models}},\ \href@noop {} {Ph.D. thesis},\ \bibinfo  {school} {California
  Institute of Technology} (\bibinfo {year} {1991})\BibitemShut {NoStop}%
\bibitem [{\citenamefont {Graczyk}\ \emph {et~al.}(2010)\citenamefont
  {Graczyk}, \citenamefont {Plonski},\ and\ \citenamefont
  {Sulej}}]{Graczyk:2010gw}%
  \BibitemOpen
  \bibfield  {author} {\bibinfo {author} {\bibfnamefont {K.~M.}\ \bibnamefont
  {Graczyk}}, \bibinfo {author} {\bibfnamefont {P.}~\bibnamefont {Plonski}}, \
  and\ \bibinfo {author} {\bibfnamefont {R.}~\bibnamefont {Sulej}},\ }\href
  {\doibase 10.1007/JHEP09(2010)053} {\bibfield  {journal} {\bibinfo  {journal}
  {JHEP}\ }\textbf {\bibinfo {volume} {09}},\ \bibinfo {pages} {053} (\bibinfo
  {year} {2010})},\ \Eprint {http://arxiv.org/abs/1006.0342} {arXiv:1006.0342
  [hep-ph]} \BibitemShut {NoStop}%
\bibitem [{\citenamefont {Graczyk}(2011)}]{Graczyk:2011kh}%
  \BibitemOpen
  \bibfield  {author} {\bibinfo {author} {\bibfnamefont {K.~M.}\ \bibnamefont
  {Graczyk}},\ }\href {\doibase 10.1103/PhysRevC.84.034314} {\bibfield
  {journal} {\bibinfo  {journal} {Phys. Rev.}\ }\textbf {\bibinfo {volume}
  {C84}},\ \bibinfo {pages} {034314} (\bibinfo {year} {2011})},\ \Eprint
  {http://arxiv.org/abs/1106.1204} {arXiv:1106.1204 [hep-ph]} \BibitemShut
  {NoStop}%
\bibitem [{\citenamefont {Graczyk}(2013)}]{Graczyk:2013pca}%
  \BibitemOpen
  \bibfield  {author} {\bibinfo {author} {\bibfnamefont {K.~M.}\ \bibnamefont
  {Graczyk}},\ }\href {\doibase 10.1103/PhysRevC.88.065205} {\bibfield
  {journal} {\bibinfo  {journal} {Phys. Rev.}\ }\textbf {\bibinfo {volume}
  {C88}},\ \bibinfo {pages} {065205} (\bibinfo {year} {2013})},\ \Eprint
  {http://arxiv.org/abs/1306.5991} {arXiv:1306.5991 [hep-ph]} \BibitemShut
  {NoStop}%
\bibitem [{\citenamefont {Graczyk}\ and\ \citenamefont
  {Juszczak}(2015{\natexlab{a}})}]{Graczyk:2014coa}%
  \BibitemOpen
  \bibfield  {author} {\bibinfo {author} {\bibfnamefont {K.~M.}\ \bibnamefont
  {Graczyk}}\ and\ \bibinfo {author} {\bibfnamefont {C.}~\bibnamefont
  {Juszczak}},\ }\href {\doibase 10.1088/0954-3899/42/3/034019} {\bibfield
  {journal} {\bibinfo  {journal} {J. Phys.}\ }\textbf {\bibinfo {volume}
  {G42}},\ \bibinfo {pages} {034019} (\bibinfo {year} {2015}{\natexlab{a}})},\
  \Eprint {http://arxiv.org/abs/1409.5244} {arXiv:1409.5244 [hep-ph]}
  \BibitemShut {NoStop}%
\bibitem [{\citenamefont {Graczyk}\ and\ \citenamefont
  {Juszczak}(2015{\natexlab{b}})}]{Graczyk:2015kka}%
  \BibitemOpen
  \bibfield  {author} {\bibinfo {author} {\bibfnamefont {K.~M.}\ \bibnamefont
  {Graczyk}}\ and\ \bibinfo {author} {\bibfnamefont {C.}~\bibnamefont
  {Juszczak}},\ }\href {\doibase 10.1103/PhysRevC.91.045205} {\bibfield
  {journal} {\bibinfo  {journal} {Phys. Rev.}\ }\textbf {\bibinfo {volume}
  {C91}},\ \bibinfo {pages} {045205} (\bibinfo {year}
  {2015}{\natexlab{b}})}\BibitemShut {NoStop}%
\bibitem [{\citenamefont {{F. Rosenblatt}}(1962)}]{Rosenblatt62}%
  \BibitemOpen
  \bibfield  {author} {\bibinfo {author} {\bibnamefont {{F. Rosenblatt}}},\
  }\href {http://www.dtic.mil/docs/citations/AD0256582} {\emph {\bibinfo
  {title} {{Principles of Neurodynamics}}}}\ (\bibinfo  {publisher} {New York:
  Spartan},\ \bibinfo {year} {1962})\BibitemShut {NoStop}%
\bibitem [{\citenamefont {Cybenko}(1989)}]{Cybenko_Theorem}%
  \BibitemOpen
  \bibfield  {author} {\bibinfo {author} {\bibfnamefont {G.}~\bibnamefont
  {Cybenko}},\ }\href {\doibase 10.1007/BF02551274} {\bibfield  {journal}
  {\bibinfo  {journal} {Math Control, Signal}\ }\textbf {\bibinfo {volume}
  {2}},\ \bibinfo {pages} {303} (\bibinfo {year} {1989})}\BibitemShut {NoStop}%
\bibitem [{\citenamefont {Funahashi}(1989)}]{FUNAHASHI1989183}%
  \BibitemOpen
  \bibfield  {author} {\bibinfo {author} {\bibfnamefont {K.-I.}\ \bibnamefont
  {Funahashi}},\ }\href {\doibase https://doi.org/10.1016/0893-6080(89)90003-8}
  {\bibfield  {journal} {\bibinfo  {journal} {Neural Networks}\ }\textbf
  {\bibinfo {volume} {2}},\ \bibinfo {pages} {183 } (\bibinfo {year}
  {1989})}\BibitemShut {NoStop}%
\bibitem [{\citenamefont {Hecht-Nielsen}(1989)}]{118638}%
  \BibitemOpen
  \bibfield  {author} {\bibinfo {author} {\bibfnamefont {R.}~\bibnamefont
  {Hecht-Nielsen}},\ }in\ \href {\doibase 10.1109/IJCNN.1989.118638} {\emph
  {\bibinfo {booktitle} {International 1989 Joint Conference on Neural
  Networks}}}\ (\bibinfo {year} {1989})\ pp.\ \bibinfo {pages} {593--605
  vol.1}\BibitemShut {NoStop}%
\bibitem [{\citenamefont {Cotter}(1990)}]{80265}%
  \BibitemOpen
  \bibfield  {author} {\bibinfo {author} {\bibfnamefont {N.~E.}\ \bibnamefont
  {Cotter}},\ }\href {\doibase 10.1109/72.80265} {\bibfield  {journal}
  {\bibinfo  {journal} {IEEE Transactions on Neural Networks}\ }\textbf
  {\bibinfo {volume} {1}},\ \bibinfo {pages} {290} (\bibinfo {year}
  {1990})}\BibitemShut {NoStop}%
\bibitem [{\citenamefont {Ito}(1991)}]{ITO1991385}%
  \BibitemOpen
  \bibfield  {author} {\bibinfo {author} {\bibfnamefont {Y.}~\bibnamefont
  {Ito}},\ }\href {\doibase https://doi.org/10.1016/0893-6080(91)90075-G}
  {\bibfield  {journal} {\bibinfo  {journal} {Neural Networks}\ }\textbf
  {\bibinfo {volume} {4}},\ \bibinfo {pages} {385 } (\bibinfo {year}
  {1991})}\BibitemShut {NoStop}%
\bibitem [{\citenamefont {Kreinovich}(1991)}]{KREINOVICH1991381}%
  \BibitemOpen
  \bibfield  {author} {\bibinfo {author} {\bibfnamefont {V.~Y.}\ \bibnamefont
  {Kreinovich}},\ }\href {\doibase
  https://doi.org/10.1016/0893-6080(91)90074-F} {\bibfield  {journal} {\bibinfo
   {journal} {Neural Networks}\ }\textbf {\bibinfo {volume} {4}},\ \bibinfo
  {pages} {381 } (\bibinfo {year} {1991})}\BibitemShut {NoStop}%
\bibitem [{\citenamefont {Geman}\ \emph {et~al.}(1992)\citenamefont {Geman},
  \citenamefont {Bienenstock},\ and\ \citenamefont
  {Doursat}}]{Geman.1992.4.1.1}%
  \BibitemOpen
  \bibfield  {author} {\bibinfo {author} {\bibfnamefont {S.}~\bibnamefont
  {Geman}}, \bibinfo {author} {\bibfnamefont {E.}~\bibnamefont {Bienenstock}},
  \ and\ \bibinfo {author} {\bibfnamefont {R.}~\bibnamefont {Doursat}},\ }\href
  {\doibase 10.1162/neco.1992.4.1.1} {\bibfield  {journal} {\bibinfo  {journal}
  {Neural Computation}\ }\textbf {\bibinfo {volume} {4}},\ \bibinfo {pages} {1}
  (\bibinfo {year} {1992})},\ \Eprint
  {http://arxiv.org/abs/https://doi.org/10.1162/neco.1992.4.1.1}
  {https://doi.org/10.1162/neco.1992.4.1.1} \BibitemShut {NoStop}%
\bibitem [{\citenamefont {Berger}\ and\ \citenamefont
  {Jefferys}(1992)}]{Berger1992}%
  \BibitemOpen
  \bibfield  {author} {\bibinfo {author} {\bibfnamefont {J.~O.}\ \bibnamefont
  {Berger}}\ and\ \bibinfo {author} {\bibfnamefont {W.~H.}\ \bibnamefont
  {Jefferys}},\ }\href {\doibase 10.1007/BF02589047} {\bibfield  {journal}
  {\bibinfo  {journal} {Journal of the Italian Statistical Society}\ }\textbf
  {\bibinfo {volume} {1}},\ \bibinfo {pages} {17} (\bibinfo {year}
  {1992})}\BibitemShut {NoStop}%
\bibitem [{\citenamefont {Jefferys}\ and\ \citenamefont
  {Berger}(1992)}]{Jefferys_bis_1992}%
  \BibitemOpen
  \bibfield  {author} {\bibinfo {author} {\bibfnamefont {W.~H.}\ \bibnamefont
  {Jefferys}}\ and\ \bibinfo {author} {\bibfnamefont {J.~O.}\ \bibnamefont
  {Berger}},\ }\href {http://www.jstor.org/stable/29774559} {\bibfield
  {journal} {\bibinfo  {journal} {American Scientist}\ }\textbf {\bibinfo
  {volume} {80}},\ \bibinfo {pages} {64} (\bibinfo {year} {1992})}\BibitemShut
  {NoStop}%
\bibitem [{\citenamefont {Pohl}\ \emph {et~al.}(2010)\citenamefont {Pohl},
  \citenamefont {Antognini}, \citenamefont {Nez}, \citenamefont {Amaro},
  \citenamefont {Biraben} \emph {et~al.}}]{Pohl:2010zza}%
  \BibitemOpen
  \bibfield  {author} {\bibinfo {author} {\bibfnamefont {R.}~\bibnamefont
  {Pohl}}, \bibinfo {author} {\bibfnamefont {A.}~\bibnamefont {Antognini}},
  \bibinfo {author} {\bibfnamefont {F.}~\bibnamefont {Nez}}, \bibinfo {author}
  {\bibfnamefont {F.~D.}\ \bibnamefont {Amaro}}, \bibinfo {author}
  {\bibfnamefont {F.}~\bibnamefont {Biraben}},  \emph {et~al.},\ }\href
  {\doibase 10.1038/nature09250} {\bibfield  {journal} {\bibinfo  {journal}
  {Nature}\ }\textbf {\bibinfo {volume} {466}},\ \bibinfo {pages} {213}
  (\bibinfo {year} {2010})}\BibitemShut {NoStop}%
\bibitem [{\citenamefont {Sick}(2018)}]{Sick:2018fzn}%
  \BibitemOpen
  \bibfield  {author} {\bibinfo {author} {\bibfnamefont {I.}~\bibnamefont
  {Sick}},\ }\href {\doibase 10.3390/atoms6010002} {\bibfield  {journal}
  {\bibinfo  {journal} {Atoms}\ }\textbf {\bibinfo {volume} {6}},\ \bibinfo
  {pages} {2} (\bibinfo {year} {2018})},\ \Eprint
  {http://arxiv.org/abs/1801.01746} {arXiv:1801.01746 [nucl-ex]} \BibitemShut
  {NoStop}%
\bibitem [{\citenamefont {MacKay}(1992{\natexlab{a}})}]{MacKay1992.4.3.415}%
  \BibitemOpen
  \bibfield  {author} {\bibinfo {author} {\bibfnamefont {D.~J.~C.}\
  \bibnamefont {MacKay}},\ }\href {\doibase 10.1162/neco.1992.4.3.415}
  {\bibfield  {journal} {\bibinfo  {journal} {Neural Computation}\ }\textbf
  {\bibinfo {volume} {4}},\ \bibinfo {pages} {415} (\bibinfo {year}
  {1992}{\natexlab{a}})},\ \Eprint
  {http://arxiv.org/abs/https://doi.org/10.1162/neco.1992.4.3.415}
  {https://doi.org/10.1162/neco.1992.4.3.415} \BibitemShut {NoStop}%
\bibitem [{\citenamefont {MacKay}(1992{\natexlab{b}})}]{MacKay1992.4.3.448}%
  \BibitemOpen
  \bibfield  {author} {\bibinfo {author} {\bibfnamefont {D.~J.~C.}\
  \bibnamefont {MacKay}},\ }\href {\doibase 10.1162/neco.1992.4.3.448}
  {\bibfield  {journal} {\bibinfo  {journal} {Neural Computation}\ }\textbf
  {\bibinfo {volume} {4}},\ \bibinfo {pages} {448} (\bibinfo {year}
  {1992}{\natexlab{b}})},\ \Eprint
  {http://arxiv.org/abs/https://doi.org/10.1162/neco.1992.4.3.448}
  {https://doi.org/10.1162/neco.1992.4.3.448} \BibitemShut {NoStop}%
\bibitem [{\citenamefont {Berger}(1985)}]{Berger_1985}%
  \BibitemOpen
  \bibfield  {author} {\bibinfo {author} {\bibfnamefont {J.~O.}\ \bibnamefont
  {Berger}},\ }\href {\doibase 10.1007/978-1-4757-4286-2} {\emph {\bibinfo
  {title} {Statistical Decision Theory and Bayesian Analysis}}}\ (\bibinfo
  {publisher} {Springer-Verlag New York},\ \bibinfo {year} {1985})\BibitemShut
  {NoStop}%
\bibitem [{\citenamefont {Llewellyn~Smith}(1972)}]{LlewellynSmith:1971uhs}%
  \BibitemOpen
  \bibfield  {author} {\bibinfo {author} {\bibfnamefont {C.~H.}\ \bibnamefont
  {Llewellyn~Smith}},\ }\bibfield  {booktitle} {\emph {\bibinfo {booktitle}
  {{Gauge Theories and Neutrino Physics, Jacob, 1978:0175}}},\ }\href {\doibase
  10.1016/0370-1573(72)90010-5} {\bibfield  {journal} {\bibinfo  {journal}
  {Phys. Rept.}\ }\textbf {\bibinfo {volume} {3}},\ \bibinfo {pages} {261}
  (\bibinfo {year} {1972})}\BibitemShut {NoStop}%
\bibitem [{\citenamefont {Galster}\ \emph {et~al.}(1971)\citenamefont
  {Galster}, \citenamefont {Klein}, \citenamefont {Moritz}, \citenamefont
  {Schmidt}, \citenamefont {Wegener},\ and\ \citenamefont
  {Bleckwenn}}]{Galster:1971kv}%
  \BibitemOpen
  \bibfield  {author} {\bibinfo {author} {\bibfnamefont {S.}~\bibnamefont
  {Galster}}, \bibinfo {author} {\bibfnamefont {H.}~\bibnamefont {Klein}},
  \bibinfo {author} {\bibfnamefont {J.}~\bibnamefont {Moritz}}, \bibinfo
  {author} {\bibfnamefont {K.~H.}\ \bibnamefont {Schmidt}}, \bibinfo {author}
  {\bibfnamefont {D.}~\bibnamefont {Wegener}}, \ and\ \bibinfo {author}
  {\bibfnamefont {J.}~\bibnamefont {Bleckwenn}},\ }\href {\doibase
  10.1016/0550-3213(71)90068-X} {\bibfield  {journal} {\bibinfo  {journal}
  {Nucl. Phys.}\ }\textbf {\bibinfo {volume} {B32}},\ \bibinfo {pages} {221}
  (\bibinfo {year} {1971})}\BibitemShut {NoStop}%
\bibitem [{\citenamefont {Nieves}\ \emph {et~al.}(2004)\citenamefont {Nieves},
  \citenamefont {Amaro},\ and\ \citenamefont {Valverde}}]{Nieves:2004wx}%
  \BibitemOpen
  \bibfield  {author} {\bibinfo {author} {\bibfnamefont {J.}~\bibnamefont
  {Nieves}}, \bibinfo {author} {\bibfnamefont {J.~E.}\ \bibnamefont {Amaro}}, \
  and\ \bibinfo {author} {\bibfnamefont {M.}~\bibnamefont {Valverde}},\ }\href
  {\doibase 10.1103/PhysRevC.70.055503, 10.1103/PhysRevC.72.019902} {\bibfield
  {journal} {\bibinfo  {journal} {Phys. Rev.}\ }\textbf {\bibinfo {volume}
  {C70}},\ \bibinfo {pages} {055503} (\bibinfo {year} {2004})},\ \bibinfo
  {note} {[Erratum: Phys. Rev.C72,019902(2005)]},\ \Eprint
  {http://arxiv.org/abs/nucl-th/0408005} {arXiv:nucl-th/0408005 [nucl-th]}
  \BibitemShut {NoStop}%
\bibitem [{\citenamefont {Singh}\ and\ \citenamefont
  {Arenhövel}(1986)}]{Singh:1986xh}%
  \BibitemOpen
  \bibfield  {author} {\bibinfo {author} {\bibfnamefont {S.~K.}\ \bibnamefont
  {Singh}}\ and\ \bibinfo {author} {\bibfnamefont {H.}~\bibnamefont
  {Arenhövel}},\ }\href {\doibase 10.1007/BF01294589} {\bibfield  {journal}
  {\bibinfo  {journal} {Z. Phys.}\ }\textbf {\bibinfo {volume} {A324}},\
  \bibinfo {pages} {347} (\bibinfo {year} {1986})}\BibitemShut {NoStop}%
\bibitem [{\citenamefont {Barish}\ \emph {et~al.}(1979)\citenamefont {Barish}
  \emph {et~al.}}]{Barish:1978pj}%
  \BibitemOpen
  \bibfield  {author} {\bibinfo {author} {\bibfnamefont {S.~J.}\ \bibnamefont
  {Barish}} \emph {et~al.},\ }\href {\doibase 10.1103/PhysRevD.19.2521}
  {\bibfield  {journal} {\bibinfo  {journal} {Phys. Rev.}\ }\textbf {\bibinfo
  {volume} {D19}},\ \bibinfo {pages} {2521} (\bibinfo {year}
  {1979})}\BibitemShut {NoStop}%
\bibitem [{\citenamefont {D'Agostini}(1994)}]{DAgostini:1993arp}%
  \BibitemOpen
  \bibfield  {author} {\bibinfo {author} {\bibfnamefont {G.}~\bibnamefont
  {D'Agostini}},\ }\href {\doibase 10.1016/0168-9002(94)90719-6} {\bibfield
  {journal} {\bibinfo  {journal} {Nucl. Instrum. Meth.}\ }\textbf {\bibinfo
  {volume} {A346}},\ \bibinfo {pages} {306} (\bibinfo {year}
  {1994})}\BibitemShut {NoStop}%
\bibitem [{\citenamefont {Graczyk}\ \emph {et~al.}(2009)\citenamefont
  {Graczyk}, \citenamefont {Kielczewska}, \citenamefont {Przewlocki},\ and\
  \citenamefont {Sobczyk}}]{Graczyk:2009qm}%
  \BibitemOpen
  \bibfield  {author} {\bibinfo {author} {\bibfnamefont {K.~M.}\ \bibnamefont
  {Graczyk}}, \bibinfo {author} {\bibfnamefont {D.}~\bibnamefont
  {Kielczewska}}, \bibinfo {author} {\bibfnamefont {P.}~\bibnamefont
  {Przewlocki}}, \ and\ \bibinfo {author} {\bibfnamefont {J.~T.}\ \bibnamefont
  {Sobczyk}},\ }\href {\doibase 10.1103/PhysRevD.80.093001} {\bibfield
  {journal} {\bibinfo  {journal} {Phys. Rev.}\ }\textbf {\bibinfo {volume}
  {D80}},\ \bibinfo {pages} {093001} (\bibinfo {year} {2009})},\ \Eprint
  {http://arxiv.org/abs/0908.2175} {arXiv:0908.2175 [hep-ph]} \BibitemShut
  {NoStop}%
\bibitem [{\citenamefont {Levenberg}(1944)}]{Levenberg44}%
  \BibitemOpen
  \bibfield  {author} {\bibinfo {author} {\bibfnamefont {K.}~\bibnamefont
  {Levenberg}},\ }\href {\doibase 10.1090/qam/10666} {\bibfield  {journal}
  {\bibinfo  {journal} {Quart. Appl. Math.}\ }\textbf {\bibinfo {volume} {2}},\
  \bibinfo {pages} {164} (\bibinfo {year} {1944})}\BibitemShut {NoStop}%
\bibitem [{\citenamefont {Marquardt}(1963)}]{Marquardt63}%
  \BibitemOpen
  \bibfield  {author} {\bibinfo {author} {\bibfnamefont {D.~W.}\ \bibnamefont
  {Marquardt}},\ }\href {\doibase 10.1137/0111030} {\bibfield  {journal}
  {\bibinfo  {journal} {J. Soc. Indust. Appl. Math.}\ }\textbf {\bibinfo
  {volume} {11}},\ \bibinfo {pages} {431–441} (\bibinfo {year}
  {1963})}\BibitemShut {NoStop}%
\bibitem [{\citenamefont {Shen}\ \emph {et~al.}(2012)\citenamefont {Shen},
  \citenamefont {Marcucci}, \citenamefont {Carlson}, \citenamefont {Gandolfi},\
  and\ \citenamefont {Schiavilla}}]{Shen:2012xz}%
  \BibitemOpen
  \bibfield  {author} {\bibinfo {author} {\bibfnamefont {G.}~\bibnamefont
  {Shen}}, \bibinfo {author} {\bibfnamefont {L.~E.}\ \bibnamefont {Marcucci}},
  \bibinfo {author} {\bibfnamefont {J.}~\bibnamefont {Carlson}}, \bibinfo
  {author} {\bibfnamefont {S.}~\bibnamefont {Gandolfi}}, \ and\ \bibinfo
  {author} {\bibfnamefont {R.}~\bibnamefont {Schiavilla}},\ }\href {\doibase
  10.1103/PhysRevC.86.035503} {\bibfield  {journal} {\bibinfo  {journal} {Phys.
  Rev.}\ }\textbf {\bibinfo {volume} {C86}},\ \bibinfo {pages} {035503}
  (\bibinfo {year} {2012})},\ \Eprint {http://arxiv.org/abs/1205.4337}
  {arXiv:1205.4337 [nucl-th]} \BibitemShut {NoStop}%
\bibitem [{\citenamefont {Moreno}\ \emph {et~al.}(2015)\citenamefont {Moreno},
  \citenamefont {Donnelly}, \citenamefont {Van~Orden},\ and\ \citenamefont
  {Ford}}]{Moreno:2015nsa}%
  \BibitemOpen
  \bibfield  {author} {\bibinfo {author} {\bibfnamefont {O.}~\bibnamefont
  {Moreno}}, \bibinfo {author} {\bibfnamefont {T.~W.}\ \bibnamefont
  {Donnelly}}, \bibinfo {author} {\bibfnamefont {J.~W.}\ \bibnamefont
  {Van~Orden}}, \ and\ \bibinfo {author} {\bibfnamefont {W.~P.}\ \bibnamefont
  {Ford}},\ }\href {\doibase 10.1103/PhysRevD.92.053006} {\bibfield  {journal}
  {\bibinfo  {journal} {Phys. Rev.}\ }\textbf {\bibinfo {volume} {D92}},\
  \bibinfo {pages} {053006} (\bibinfo {year} {2015})},\ \Eprint
  {http://arxiv.org/abs/1508.00492} {arXiv:1508.00492 [hep-ph]} \BibitemShut
  {NoStop}%
\bibitem [{Wro()}]{Wroclaw}%
  \BibitemOpen
  \href@noop {} {}\bibinfo {howpublished}
  {\url{http://www.wcss.wroc.pl}}\BibitemShut {NoStop}%
\bibitem [{\citenamefont {Lepage}\ \emph {et~al.}(2002)\citenamefont {Lepage},
  \citenamefont {Clark}, \citenamefont {Davies}, \citenamefont {Hornbostel},
  \citenamefont {Mackenzie}, \citenamefont {Morningstar},\ and\ \citenamefont
  {Trottier}}]{Lepage:2001ym}%
  \BibitemOpen
  \bibfield  {author} {\bibinfo {author} {\bibfnamefont {G.~P.}\ \bibnamefont
  {Lepage}}, \bibinfo {author} {\bibfnamefont {B.}~\bibnamefont {Clark}},
  \bibinfo {author} {\bibfnamefont {C.~T.~H.}\ \bibnamefont {Davies}}, \bibinfo
  {author} {\bibfnamefont {K.}~\bibnamefont {Hornbostel}}, \bibinfo {author}
  {\bibfnamefont {P.~B.}\ \bibnamefont {Mackenzie}}, \bibinfo {author}
  {\bibfnamefont {C.}~\bibnamefont {Morningstar}}, \ and\ \bibinfo {author}
  {\bibfnamefont {H.}~\bibnamefont {Trottier}},\ }\bibfield  {booktitle} {\emph
  {\bibinfo {booktitle} {{Lattice field theory. Proceedings, 19th International
  Symposium, Lattice 2001, Berlin, Germany, August 19-24, 2001}}},\ }\href
  {\doibase 10.1016/S0920-5632(01)01638-3} {\bibfield  {journal} {\bibinfo
  {journal} {Nucl. Phys. Proc. Suppl.}\ }\textbf {\bibinfo {volume} {106}},\
  \bibinfo {pages} {12} (\bibinfo {year} {2002})},\ \Eprint
  {http://arxiv.org/abs/hep-lat/0110175} {arXiv:hep-lat/0110175 [hep-lat]}
  \BibitemShut {NoStop}%
\bibitem [{\citenamefont {Hinton}(1987)}]{10.1007/3-540-17943-7_117}%
  \BibitemOpen
  \bibfield  {author} {\bibinfo {author} {\bibfnamefont {G.~E.}\ \bibnamefont
  {Hinton}},\ }in\ \href@noop {} {\emph {\bibinfo {booktitle} {PARLE Parallel
  Architectures and Languages Europe}}},\ \bibinfo {editor} {edited by\
  \bibinfo {editor} {\bibfnamefont {J.~W.}\ \bibnamefont {de~Bakker}}, \bibinfo
  {editor} {\bibfnamefont {A.~J.}\ \bibnamefont {Nijman}}, \ and\ \bibinfo
  {editor} {\bibfnamefont {P.~C.}\ \bibnamefont {Treleaven}}}\ (\bibinfo
  {publisher} {Springer Berlin Heidelberg},\ \bibinfo {address} {Berlin,
  Heidelberg},\ \bibinfo {year} {1987})\ pp.\ \bibinfo {pages}
  {1--13}\BibitemShut {NoStop}%
\end{thebibliography}%

\end{document}